# IlocA:

## An algorithm to Cluster Cells and form Imputation Groups from a pair of Classification Variables

Gerard Keogh

Central Statistics Office, Dublin, Ireland

gerard.keogh@cso.ie

Abstract

We set out the novel bottom-up procedure to aggregate/cluster cells with small frequency counts together, in a two-way classification while maintaining dependence in the table. The procedure is model free. It combines cells in a table into clusters based on independent log-odds ratios.

We use this procedure to build a set of statistically efficient and robust imputation cells, for the imputation of missing values of a continuous variable using a pair classification variables. A nice feature of the procedure is it forms aggregation groups homogeneous with respect to the cell response mean. Using a series of simulation studies, we show IlocA only groups together *independent cells and does so* in a consistent and credible way. While imputing missing data, we show IlocA's generates close to an optimal number of imputation cells. For ignorable non-response the resulting imputed means are accurate in general. With non-ignorable missingness results are consistent with those obtained elsewhere.

We close with a case study applying our method to imputing missing building energy performance data

Keywords

## 1. Introduction

### 1.1 Motivation

Item non-response is a common feature occurring with response variables on survey datasets. It gives rise to non-response bias, which happens when respondents and non-respondents differ with respect to survey variables. Also, since the number of responses is less than the total number of dataset observations, non-response also gives rise to estimators with larger variance – this effect is called non-response variance. The main purpose of imputation is to reduce non-response bias and to control the non-response variance. Typically, this is achieved by employing one or more fully observed auxiliary variables on the survey dataset. These are chosen because they meaningfully explain both the level and variability of the response variable. In this article, we focus on the process of imputing item nonresponse using cell mean imputation, where imputation cells are constructed from a pair of fully observed auxiliary classification variables on the survey dataset. Our setting is a general-purpose survey dataset in the realm of official statistics, though our approach extends beyond this realm. At the outset, we state this is not an article about imputation methods (*e.g.,* hot-deck), but rather about a tool to create imputation cells or groups. In principle, cell mean imputation is a relatively simple imputation method. Complexity arises when we come to choose a specific set of cells that explain the response variable, from among the many possible cell permutations. Some ways to make this choice have been described in Haziza et. al. (2001), Creel & Krotki (2006) and Haziza, & Beaumont (2007).

In cell mean imputation, the survey dataset is partitioned into mutually exclusive and exhaustive subsets of observations called imputation cells (usually referred to simply as *cells*). Within each cell, the mean of the non-missing values of the response variable is computed, and this estimate is assigned independently to all missing values of that variable. In this article, we assume the response variable is continuous. To create the cells, a pair of fully observed relevant auxiliary classification variables on the survey dataset is chosen and cross-classified. In tandem, the cross-classification generates a contingency table that records the frequency count of observations falling into each cell *(i.e.,* the cell size) and the dependence between that pair of classification variables - this (tabular) dependence or association is often tested using the $\chi^2$-statistic. Our purpose is: first, to create a set of imputation cells, which have enough observations to ensure the overall response variable mean is statistically efficient and robust (meaning, less sensitive to cell misspecification); and second, ensure imputed missing values respect the dependence that exists between the pair of auxiliary classification variables. Following Ballin & Barcaroli (2013), we refer to the cells defined by the full two-way cross-classification of the chosen pair of auxiliary variable classes, as *atomic cells*. Meanwhile, we refer to the case where auxiliary classifications are completely ignored and the dataset is not partitioned in any way, as the *null cell*.

Kalton & Kasprzyk (1986) highlight two limitations associated with cell (or in their terminology, class) mean imputation. First, when atomic cells are employed, the number of cells may be quite large and consequently, the cell size of several cells may be small. Consequently, cell mean estimates will not be statistically efficient due to their small size, nor will they be robust to small changes in the number of cells. According to Haziza et. al (2001), one straightforward way to mitigate these problems is by aggregating (or merging) smaller atomic cells into a larger *aggregation group* of cells. Second, cell mean imputation can have harmful effects on the analyses of relationships, often attenuating associations between variables. Clearly, once we aggregate cells, we interfere with the association (dependence) between the auxiliary variables that exists at the atomic cell level. One way to mitigate this and maintain dependent relationships, is to only aggregate atomic cells that are independent. The resulting aggregation groups will respect dependent relationships and their response variable group means will be statistically more efficient and robust.

To address the limitations of cell mean imputation described above, we set out a novel and relatively simple algorithmic procedure to create aggregation groups. The procedure is called IlocA (Independent log-odds cell Aggregation). Compute code to implement the procedure is provided in the Appendix. The method takes the contingency table generated from the pair of auxiliary classification variables, and forms aggregation groups of independent atomic cells in a *bottom-up* manner. It merges atomic cells with a small frequency count, say less than 20, into a larger aggregation groups until a pre-set overall aggregation group frequency is exceeded. Only atomic cells within the contingency table that contribute to *close to zero* log-odds ratios are aggregated. An odds (or cross product) ratio is the ratio of the product of the two diagonal atomic cell values, to the product of the two off-diagonal atomic cell values, in a two-by-two sub-table. For example, in a two-by-two table $\begin{bmatrix} 5 & 3 \\ 2 & 4 \end{bmatrix}$ the odds ratio is $3\frac{1}{3} = (5 \times 4) \div (3 \times 2)$.

A log-odds ratio is the log of the odds ratio and a value of zero defines independence effects (see Agresti, 2002). By only aggregating cells used to compute independent log-odds ratios, our procedure strives to maintain association between the two auxiliary classification variables defining the contingency table. Applying cell mean imputation within the imputation cells defined by the resulting aggregation groups, ensures the imputed values respect the dependence between the pair of auxiliary classification variables. Meanwhile, by creating aggregation groups that are large, IlocA ensures the overall response variable mean is statistically more efficient and robust. Note, our procedure makes no direct attempt to aggregate the response variable itself into homogenous response variable strata.

**1.2 Some alternative imputation strategies**

For our analysis, we favour cell mean imputation as it is one of the simplest and most widely understood imputation methods available. Other, more complex procedures are described in Kalton & Kasprzyk (1986), or alternatively Fractional Hot Deck Imputation (Kim & Fuller, 2004) may be employed. However, many of these methods are far less intuitive and require a high level of sophistication to apply.

Ideally, we should strive to form groups of cells where the response values within each group are a similar size - these are called homogenous response variable groups, or more commonly homogenous groups. The idea of forming homogeneous aggregation groups in the response variable has a long history in the realm of stratified sampling, going back to the $cum\sqrt{f}$ method of Dalineus and Hodges (1959), see also Cochran (1977, Ch 5), and in the multi-response case to Bethel (1989). The $cum\sqrt{f}$ method is a quantile method that is often applied to tabulated response data. Beyond that realm Moses et. al. (1969) and Bloch & Segal (1989) applied the $cum\sqrt{f}$ method to define so-called *pure-aggregation strata*. IlocA fits within this *canon* in that it too aggregates cross-tabulated data. However, IlocA does not form homogenous response variable aggregation strata. Rather, it takes the atomic cells in the contingency table created from the cross-classification of a pair of auxiliary classification variables and forms those cells into heterogenous clusters based on cell size. These clusters are formed so that they maintain the original two-way tabular association (dependence) that existed between the atomic cells in the contingency table. Clearly, in the imputation setting, focussing solely on the auxiliary variables while ignoring the response variable with its missing data, is a bad idea. But of course, auxiliary variables are chosen because they explain the response variable. Indeed, IlocA also incorporates cell means and standard deviations in to process of building aggregation groups. On this basis, IlocA will preserve response variable homogeneity for those response values that fall in atomic cells that are not aggregated *(i.e.,* dependent cells). Meanwhile, within aggregated cells, response variable homogeneity is attenuated in order to provide sufficient responses to compute statistically efficient and robust within aggregation/imputation cell mean estimates.

Of course, a bottom-up approach is not the only way to proceed to form larger aggregation groups of atomic cells. Parametrically, one obvious approach that takes account of two-way relationships involves fitting a log-linear model to the contingency table, followed by aggregating cells where association terms are not significant. More generally, a linear model may be fitted to the response variable itself using a set of auxiliary predictor variables, followed by aggregation. In this vein Haziza et. al. (2001) describe an elegant method that combines quantile aggregation and regression. It builds aggregation groups from quantiles of the predicted response values and/or response probabilities. Several non-parametric procedures may also be employed. Simple non-parametric procedures include k-Means or k-Nearest Neighbours. More sophisticated procedures include *top-down* recursive tree partitioning procedures such as CART (Classification and Regression Trees, Breiman et. al. 1984) or CHAID (Kass, 1980). Steinberg & Colla (1995), for example, have used CART to form imputation cells. Model based clustering methods, such as 'mclust' within R might also be considered. We have some experience of using several of these methods to partition a two-way contingency table and found they gave mixed results. For example, the $cum\sqrt{f}$ method produced aggregation groups with reasonable sized frequency counts, but of course, those groups cannot maintain table relationships. Using loglinear models, we tended to find many model interaction terms were statistically significant, but had little impact. Unlike Creel & Krotki (2006), we found CART, running in R, gave quite large partitions that were hard to interpret – though, we note Creel & Krotki restricted the tree to a depth of three. CHAID (HPSPLIT procedure in SAS), on the other hand, gave sensible aggregation groups of cells, but on occasion produced some groups with extremely large frequency counts. For CHAID to be practical, it would be nice to be able to set minimum and maximums bounds on the size of the aggregation group of cells it generates. Unfortunately, this facility is not available.

**1.3 Geographic tables, missingness and article structure**

Interestingly, IlocA was initially conceived for use on a survey dataset where one of the auxiliary classification variables recorded geographic population data. Typically, this gives rise to a two-way contingency table, where there is one small block of atomic cells with large frequency counts, say relating to cities, and another large block of mainly small cell counts usually relating to rural counties. In this instance, there is a good chance of finding near-zero log-odds ratios, upon which independent aggregation groups within the contingency table may be formed. IlocA was conceived to exploit this feature of geographic tables. We explore the consequences of this further in Section 2.

In practice the missing data mechanism is of particular interest. We assume the missing data mechanism of our response variable is missing at random. This means the distribution of the response variable is the same for both response and missing values when we control on an auxiliary variable or factor. Accordingly, the statistical process generating the missing data – the so-called missing data mechanism – is ignorable (see Haziza et. al. 2001) within atomic cells constructed using the auxiliary variables, which of course are chosen *a priori* because they explain both the level and variability of the response variable. Having said that, we also explore in simulation studies the performance of IlocA when the missing data mechanism in not-ignorable and here we find IlocA competes well with other methods.

Section 2 provides some relevant background on log-odds ratios, including a statement that the density of the log-odds ratios of a blocked contingency table has dominant central peak. Section 3 sets out the IlocA procedure. In section 4 we conduct a series of simulation studies, to see whether IlocA performs credibly as a conventional statistical table clustering method. Section 5 describes the performance of IlocA when used to construct a set of imputation cells from auxiliary classification variables. We show the set of aggregated cells generated by IlocA, gives efficient estimates of the overall dataset mean value under two data generating models and across two distinct non-response models. Correctly specified and mis-specified non-response models are considered as well as different levels of non-response. In section 6 we report on a case study. Here, IlocM is used to incorporate a geographic effect into the imputation of dwelling energy performance values. Section 7 concludes.

## 2. Two-way and blocked two-way contingency table log-odds-ratios

For a two-way contingency table of frequency counts $A$, of dimension $m \times n$, let $a_{ij}$ denote the value in cell $(i,j)$ as follows:

$$A = \begin{bmatrix} a_{11} & a_{12} & & & \cdots & & a_{1n} \\ & & a_{ij} & \cdots & a_{il} & & \\ \vdots & & \vdots & & \vdots & \vdots & \\ & & a_{kj} & \cdots & a_{kl} & & \\ & & & \cdots & & & \\ a_{m1} & & & & & & a_{mn} \end{bmatrix} \begin{matrix} R_1 \\ R_i \\ \vdots \\ R_k \\ \\ R_m \end{matrix}$$
$$\quad\; C_1 \quad\;\; C_j \;\cdots\; C_l \;\; C_n \quad T$$

Row and column totals are denoted by $R$ and $C$ respectively while the overall total is labelled $T$. On table $A$, it is convenient to identify a two-by-two sub-table labelled $A_{2\times 2}$:

$$A_{2\times 2}(ijkl) = \begin{bmatrix} a_{ij} & a_{il} \\ a_{kj} & a_{kl} \end{bmatrix}$$

For table $A$, the multiplicative/loglinear model relates the cell value $a_{ij}$ to: the overall total, row and column totals, a within cell interaction effect $g_{ij}$ and a residual cell effect $e_{ij}$ according to

$$a_{ij} = T \times R_i \times C_j \times g_{ij} \times e_{ij} = \exp(\tau + \alpha_i + \beta_j + \gamma_{ij}) \times \exp(\varepsilon_{ij})$$
$$\Leftrightarrow \quad \log(a_{ij}) = \tau + \alpha_i + \beta_j + \gamma_{ij} + \varepsilon_{ij}$$
(1)

where $T = e^\tau$, $R_i = e^{\alpha_i}$ and $C_j = e^{\beta_j}$, $g_{ij} = e^{\gamma_{ij}}$ and $e_{ij} = e^{\varepsilon_{ij}}$ and $\tau, \alpha_i, \beta_j, \gamma_{ij}$ and $\varepsilon_{ij}$ are the overall total, row, column, interaction and residual effects on the log scale respectively. When we assume the actual cell counts $a_{ij}$ are Poisson random variables with mean (expected) values $\bar{a}_{ij} = \exp(\tau + \alpha_i + \beta_j + \gamma_{ij})$, it is standard to incorporate the randomness directly into the statistical model and write

$$a_{ij} \sim Po(\bar{a}_{ij})$$

This is the conventional Poisson-Multinomial model for the expected counts, as described in Agresti (2002, Ch. 5 & 6). The parameters of the loglinear model in this form are estimated by standard software packages (*e.g.,* Proc CATMOD/GENMOD in SAS). That said, we find it convenient to assume the statistical model is lognormal. The randomness in the log residuals $\varepsilon_{ij}$ in (1) is therefore described by a $N(0, \sigma^2)$ distribution – we justify this assumption below.

Association, or two-way interaction between cells in the table may be described using cross-product ratios. In the realm of applied statistics, the cross-product ratio is commonly referred to as the odds ratio $\theta_{ijkl}$. Using (1) the odds ratio and concomitant log-odds ratio $\phi_{ijkl}$ for the two-by-two sub-table $A_{2\times 2}(ijkl)$ shown above, are then given by the formulae

$$\theta_{ijkl} = \frac{a_{ij}a_{kl}}{a_{il}a_{kj}} = \frac{\exp(\gamma_{ij}+\gamma_{kl}) \times \exp(\varepsilon_{ij}+\varepsilon_{kl})}{\exp(\gamma_{il}+\gamma_{kj}) \times \exp(\varepsilon_{il}+\varepsilon_{kj})}$$

$$\phi_{ijkl} = \log(\theta_{ijkl}) = \begin{cases} \log(a_{ij}) + \log(a_{kl}) - \log(a_{il}) - \log(a_{kj}) \\ \gamma_{ij} + \gamma_{kl} - \gamma_{il} - \gamma_{kj} + \eta_{ijkl} \end{cases} \quad (2)$$

Under our assumption $\varepsilon_{ij} \sim N(0, \sigma^2)$, we then have $\eta_{ijkl} = \varepsilon_{ij} + \varepsilon - \varepsilon_{il} - \varepsilon_{kj} \sim N(0, 4\sigma^2)$.

A number points are worth highlighting here. First, when all $a_{ij} > 0$ there are $m(m-1) \times n(n-1)/4$ unique odds/log-odds ratios associated with $A$. If, in addition, an $a_{ij} = 0$, or is a repeated value, this number is reduced by $(m-1) \times (n-1)$. Moreover, when the rows and columns of $A$ are independent (*i.e.*, we have an independence table) then $\theta_{ijkl} = 1$, since $\theta_{ijkl} = \frac{a_{ij}a_{kl}}{a_{il}a_{kj}} = \frac{R_iC_j \times R_kC_l}{R_iC_l \times R_kC_j} = 1$ and so $\phi_{ijkl} = 0$.

Second, under the conventional Poisson sampling model for the cell value, a formal representation for the distribution of the log-odds ratios $\phi_{ijkl}$, is generally not readily available. In this case the distribution of the log-odds ratios is complex, as it is built up from log zero-truncated Poisson distributions. Accordingly, the likelihood function for the log-odds ratios cannot be directly written down. In contrast, Agresti (2002), Ch 6 constructs the likelihood for the cell values and treats the log-odds ratio as functions of these quantities. The distribution of the log-odds ratios is then treated from an asymptotic perspective only.

Third, equation (2) shows the log-odds ratios $\phi_{ijkl}$ describe the association between cell values in terms of table interaction effects.

Fourth, under this Normal model for the log table values, the likelihood function for the $\eta_{ijkl}$ parameters is

$$L = \prod_{ijkl} \frac{1}{\sqrt{4\pi\sigma}} \exp^{-\frac{1}{2}\left(\frac{\eta_{ijkl}}{2\sigma}\right)^2}$$

This likelihood function contrasts significantly with the Poisson form given in Agresti (2002, Ch 6). Nonetheless, the resulting association parameter estimates should be close for a large table.

Fifth, from (2) the distribution of the log-odds ratios $\phi_{ijkl} \sim N(0, 4\sigma^2)$ since $\eta_{ijkl} \sim N(0, 4\sigma^2)$. According to Powers and Xie (2008), a Normal distribution of the log-odds ratios is frequently observed in practice. So, on this basis the normality assumption of the log $a_{ij}$ values make sense. To the best of our knowledge, the phenomenon whereby the log-odds ratios in a large contingency table tend to be normally distributed, was first identified by Good (1956, 1965 p59). He suggested the effect begins to become manifest in $5 \times 5$ tables. In any event, in a large table we might expect $\sigma^2$, the variance of the log $a_{ij}$ values, to be quite large. On this basis the variance of the log-odds ratios at $4\sigma^2$ is very large, so their distribution will quite flat. Consequently, the relative probability of finding independence in a large table is small, since the probability of near-zero log-odds ratios is relatively small. This suggests the scope to aggregate cells in a large contingency table, while preserving dependence may be somewhat limited as the table size increases. In contrast, an under-dispersed log-odds ratio distribution will have a much higher probability of near zero log-odds. Consequently, there will be a far better chance of finding independence among cells in table where the log-odds ratio distribution is under-dispersed. This phenomenon tends to occur in geographic two-way contingency tables. On a geographic a table, we often observe there is one small block of large cell counts relating to large urban areas (*e.g.,* cities and towns), and another large block of mainly small cell counts, relating to rural areas. Accordingly, the table may be decomposed into approximate sub-blocks. These blocks define a partial ordering of the table. Therefore, assume we observe our table may be roughly split vertically (column-wise) into two vertical sub-table blocks, $A_L$ and $A_R$, with the bulk of smaller cell values predominant in $A_L$. Now, for this left and right block arrangement of the contingency table, the density of the log-odds ratios will tend to have a dominant central peak. This means the log-odds ratio distribution will be under-dispersed. Accordingly, there will be a high probability of finding near-zero log-odds ratios and independent subsets of cells within the table. We state this more formally as follows:

**Conjecture**: For a large contingency table $A$ comprising two blocks $A_L$ and $A_R$, with column dimensions $n_L$ and $n_R$ respectively, and with $n_L \gg n_R$ and $a_{ij,R} \gg a_{ij,L}$, the asymptotic density of the log-odds ratios $\phi_{ijkl}$ of $A$ has dominant central peak originating from sub-table $A_L$.

A proof of this conjecture is given in the Appendix.

Considering this, there is a group of cells with similar values within one block (*e.g.,* small values in the left block) that we can *interchange* without substantially altering the log-odds ratio distribution. It is this feature that informs our IlocA procedure which is described in the next section.

## 3. Independent log-odds cell Aggregation (IlocA) procedure

This section describes IlocA, a bottom-up procedure that aggregates/merges individual atomic cells in a contingency table into larger aggregation/cluster groups of cells based on independent log-odds ratios.

**IlocA algorithm**

> 0. Starting with a two-way contingency table $A$ having dimensions $m \times n$ and with cell counts $a_{ij}$
> 1. Set a minimum cell size $M$, maximum steps ($MS=20$ say) and $k$-proportion
> 2. While (steps $\leq$ MS and $0 < a_{ij} < M$)
>
>     2.1. Using equation (2) compute the set of log-odds ratios $\{\phi_{ijkl}\}$ of $A$.
>
>     2.2. Sort $\{\phi_{ijkl}\}$ smallest to largest and select smallest element of $\{\phi_{ijkl}\} = \phi_{ijkl,1}$
>
>     The four cell values contributing to $\phi^1_{ijkl,1}$ are $a_{ij,1}, a_{ik,1}, a_{jl,1}, a_{kl,1}$
>
>     Select the smallest of these four, such that $a_{**,1} < M$ (say, it is the first $a_{ij,1}$)
>
>     2.2.1. Set $\Sigma = a_{ij,1}$
>
>     2.2.2. Starting with the second smallest element of $\{\phi_{ijkl}\}$
>     Repeat $\tau = 2:k$ or until $\Sigma > M$
>     Compute
>     $$\delta_\tau = \max \begin{Bmatrix} |a_{ij,\tau} - a_{ij,1}|, |a_{ik,\tau} - a_{ik,1}|, \\ |a_{jl,\tau} - a_{jl,1}|, |a_{kl,\tau} - a_{kl,1}| \end{Bmatrix}$$
>
>     2.2.3. Sort $\delta_\tau$ smallest to largest and select $\tau^* = \tau$ such that $\delta_{\tau^*} = \min_\tau\{\delta_\tau\}$
>
>     2.2.4. Set (colour) the contingency table cell value $a_{ij,\tau^*} = -\text{steps}$
>
>     2.2.5. $\Sigma = \Sigma + a_{ij,\tau^*}$
>
>     2.3. Steps+1
>
>     2.4. If $\Sigma > M$ and steps $> 1$ then Set (colour) the table cell $a_{ij,1} = -\text{steps}$
> 3. STOP

The essential elements of this procedure involve, first, computing all log-odds ratios for all positive frequency counts in the contingency table and selecting the smallest log-odds ratio. We call this smallest log-odds ratio $\phi_{ijkl,1}$, the generator log-odds ratio. We then choose the minimum of the four cell values $a_{ij,1}, a_{ik,1}, a_{jl,1}, a_{kl,1}$, used to compute the generator log-odds ratio. We call this cell value, taken in this instance to be the first cell $a^1_{ij}$, the generator cell of the aggregation/cluster group of atomic cells. Then, starting with the second smallest log-odds ratio $\phi_{ijkl,2}$ in $\{\phi_{ijkl}\}$, for all *relevant* (see below) $\phi_{ijkl}$ we compute the cell distance measure $\delta_\tau$. This is the maximum absolute difference between the cell values contributing to the $\tau^{th}$ log-odds ratio, and cell values of the generator (*i.e.* first) log-odds ratio $\tau^1$. The resulting set of maximum absolute differences are then sorted smallest to largest and the first, that is, smallest $\delta_\tau$ selected. Accordingly, the optimisation criterion underpinning IlocA is mini-max – the maximum absolute difference in cell values is minimised across all log-odds ratios defining independence in the table. This process identifies the cell $a_{ij,\tau^*}$ which is assigned to a cluster/aggregation group. When the aggregation group is complete, that is when $\Sigma > M$, the procedure returns to the beginning of step 2, whereupon a new set of log-odds ratios for the table is computed using only those remaining atomic cells that have a positive frequency count.

Step 1 of the IlocA procedure sets the user-defined parameters. In practice, IlocA seems to work well with the maximum number of steps left at 20. Thus, the only input parameters required is the maximum cell size condition and the bandwidth parameter $k$-proportion. Clearly, the maximum cell size condition is set by the user according to their specific needs, though in most of our tests we have used a value of 20. We adopted this value, as it is generally considered to be a reasonable sample size figure when estimating a population mean. The bandwidth parameter $k$-proportion is used to define $k$, the number of *relevant* log-odds ratios to be included in the mini-max optimisation subset. We set $k$-proportion value equal to 1-in-20, so that the optimisation is only undertaken over the smallest 5% of log-odds ratios in absolute value. Increasing this proportion will yield a larger subset of log-odds to include in the mini-max search. However, widening the scope of the search increases the risk of including large log-odds ratios associated with dependent relationships in the contingency table. Of course, as the procedure progresses the population of log-odds ratios decreases. This means the size of the optimisation set diminishes. To prevent this, we apply a relaxation parameter to the cell size parameter $M$ - specifically, on each outer loop (step 2) of IlocA, we adjust $M$ as follows

$$M = M \times \frac{m \times n}{\left(\#\left(a_{ij} > 0\right)\right) + 1 + \text{isteps}}$$

This gradually increases the cell size criterion, so step 2 takes a while longer to terminate.

It is clear the IlocA algorithm makes no reference to response data. The only data required by the procedure, is a contingency table of frequency counts of those observations that fall into the cross-classification defined by a pair of relevant classification variables. As formulated, IlocA is therefore a contingency table cell clustering technique. Of course, when that pair of classification variables provide us with a good explanation for the response, it makes sense to consider IlocA as method to define a set of aggregation cells for imputation purposes – *i.e*, imputation cells. However, in this setting, relying solely on the cell frequency counts ignores valuable non-missing response information in the form of atomic cell means and standard deviations. Accordingly, we leverage this information by augmenting the distance measure $\delta_\tau$ to include the cell mean $\mu_{ij}$ and cell standard deviation $\sigma_{ij}$. Specifically, in step 2.2.2 we further compute absolute differences of the cell means and standard deviations according to

$$\delta_\tau(\mu) = \max \begin{cases} |\mu_{ij,\tau} - \mu_{ij,1}|, |\mu_{ik,\tau} - \mu_{ik,1}|, \\ |\mu_{jl,\tau} - \mu_{jl,1}|, |\mu_{kl,\tau} - \mu_{kl,1}| \end{cases}$$

$$\delta_\tau(\sigma) = \max \begin{cases} |\sigma_{ij,\tau} - \sigma_{ij,1}|, |\sigma_{ik,\tau} - \sigma_{ik,1}|, \\ |\sigma_{jl,\tau} - \sigma_{jl,1}|, |\sigma_{kl,\tau} - \sigma_{kl,1}| \end{cases}$$

and scale the set of each of these to lie in $[0, 1]$. Labelling the existing distance measure based on the cell frequency count $\delta_\tau(a)$, in the imputation setting we re-define

$$\delta_\tau = \delta_\tau(a) + \delta_\tau(\mu) + \delta_\tau(\sigma)$$

and aggregate atomic cells into imputation cells based on this imputation cell distance measure. Clearly, this measure uses more of the available information to arrive at a set of imputation cells and therefore provides more efficient estimates. Indeed, this imputation cell distance measure goes some way to ensure response variable homogeneity in mean (and standard deviation) is maintained. Moreover, given the one-to one correspondence between cell mean imputation and regression imputation based on indicator variables defining those cells, there is a question relating to the whether the imputation cell means, are in fact estimable functions of the actual atomic cell means. Clearly, this could be checked by computing the pseudo-inverse of the covariance matrix of the Normal equations (see Searle 1966) generated when the next atomic cell is added to the current imputation cell, say prior to the cell colouring/assignment step 2.2.4. However, we elect not to implement such a check on efficiency and simplicity grounds. Indeed, it will be clear from our simulation tests in section 5 that such a check seems unnecessary. Thus, in practice, IlocA computes imputation cell mean estimates that are themselves estimable functions of the atomic cell means.

Finally, an interesting feature of the (basic) frequency count based version of IlocA is, the minimum $\delta_\tau$ value identifies a log-odds ratio with a cell value $a_{ij,\tau^*}$, that is *close* to one of the cell values defining the generator log-odds ratio $\tau^1$. We can therefore interchange that cell value with $a_{ij,1}$ in the computation of the generator log-odds ratio $\phi_{ijkl,1}$ and get a new log-odds ratio $\phi^*_{ijkl}$ that is close to $\phi_{ijkl,1}$. More precisely,

$$P(|\phi_{ijkl,1} - \phi^*_{ijkl}| > \delta) < \epsilon \tag{3}$$

when we interchange the cell values $a_{ij,\tau^*}$ with $a_{ij,1}$. Interestingly, for blocked contingency tables this feature is more evident as most cells are in the left sub-table $A_L$, so $|a_{ij,\tau^*} - a_{ij,1}|$ is generally small. Consequently, $|\phi_{ijkl,1} - \phi^*_{ijkl}|$ is also small, so typically (3) is empirically satisfied in practice. On this basis, step 2.2.4 forms the sequence of unique interchangeable

cells into a cluster/aggregation group. Proceeding in this manner, the algorithm forms mutually exclusive and exhaustive clusters of cells in the contingency table using cells associated with independent log-odds ratios.

## 4. Empirical Simulation Study of IlocA for Cell Clustering

We first examine the performance of IlocA in conventional cell clustering setting. Accordingly, we conduct an empirical study to see whether the proposed procedure produces a set of aggregated cells that make sense. Our study consists of two parts. First, using a makeshift metric, we attempt to estimate the extent to which associations in an aggregated (cell) contingency table, are preserved when IlocA is employed. Second, we examine whether IlocA *only aggregates/merges independent cells*. Indeed, if this is the case, then cells in the table that are associated will not be aggregated and so the procedure will control for type 1 error – the possibility of aggregating cells that are associated in the table when these cells should not be aggregated. Equivalently, if there is an association between cells in the table, the procedure retains that association.

We use a *hack* of the Altham-Index (see Altham, 1970; metric(iv)) to estimate the extent to which associations in a contingency table are preserved when IlocA is employed. For an $r$-by-$c$ contingency table $A$, the independence variant of the Altham-Index is given by

$$\mathcal{L}(A) = \sum^{n_\phi} (\phi_{ijkl})^2 \qquad (4)$$

where $n_\phi$ is the number of log-odds ratios in table $A$. Clearly, contingency table frequencies associated with larger log-odds ratios will give rise to larger values of $\mathcal{L}(A) > 0$, indicating more dependence between the classification variables. The sampling distribution of this index is unknown, though for large contingency tables it may be approximated by a $\chi^2$ distribution.

Unfortunately, for IlocA's aggregated table, which we label $A^\dagger$, we cannot compute $\mathcal{L}(A^\dagger)$. However, we circumvent this difficulty by adopting a pragmatic *hack* of the Altham Index. We compute the average of frequencies in the original contingency table $A$, corresponding to each specific cluster of aggregated cells in table $A^\dagger$. We then augment merged table $A^\dagger$ by applying that average to each cell within that cluster.

We simulate data for our tests along the lines given in Kass (1980). A two-way contingency table of frequency counts of dimension $m \times n$, is generated from a dataset of $t = 1 - 480$ observations. Each observation comprises two variables $z_{1t}$ and $z_{2t}$. On each observation a random number labelled $u_t$ is drawn from the (standard) Uniform distribution. The value of variable $z_{1t}$ is cyclic and given by $z_{1t} = \text{mod}(t, m)$, while variable $z_{2t}$ is random and given by $z_{2t} = \text{int}(1 + n \times u_t)$. The contingency table is generated from the cross-classification of variable $z_{1t}$ with variable $z_{2t}$ on the dataset. We use the standard $\chi^2(m - 1 \times n - 1)$ test to check the resulting table is an independence table. IlocA is then applied to this contingency table to generate a merged table and this whole procedure is repeated 100 times. For 95 of the 100 simulated tables, we note the $\chi^2$-test is not significant, so these simulated test tables are independence tables. For comparison, we also consider a second simulation which is identical to the first, except the second variable is generated according to $z_{2t} = \text{int}(mod(1 + u_t, n))$ with $u_t \sim \text{lognormal}(0,2)$. Interestingly, tables generated by this process have one column whose cell values are generally much larger than those of the other columns, mimicking the blocking that can occur in a geographic table.

Figure 1: $\mathcal{L}(A)$ computed for each 8-by-5 table generated from each simulated dataset.

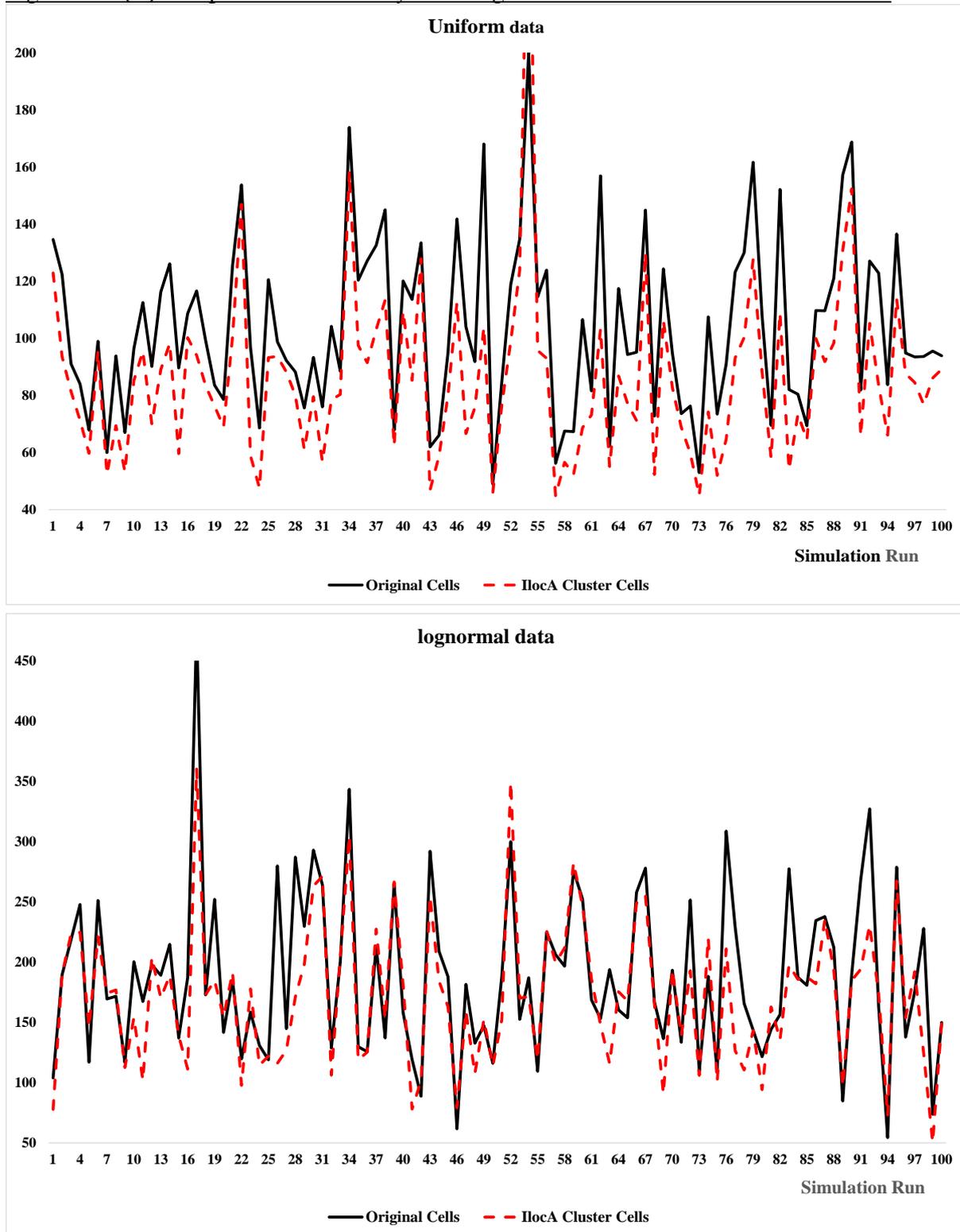

For Uniform and lognormal based tables, the value of $\mathcal{L}(A)$ is plotted in Figure 1 across all 100 simulated tables. We compare this quantity, for both the original cell and aggregated cell structures, to study the extent to which dependence is affected by adopting the IlocA's cell aggregation structure.

Examining Figure 1, it is clear the solid black line plot of $\mathcal{L}(A)$ for the original cell structure of table $A$, is generally above the dotted red line plot of $\mathcal{L}(A^\dagger)$ representing the aggregated cell structure of table $A^\dagger$. The original cell classification therefore exhibits more dependence. For this Uniform data, the average level of the black line is 104 while it is 86 for the red line. For the lognormal data the corresponding average levels are 188 and 171 respectively. For our Uniform test data, IlocA's merged cell structure therefore retains 83% of the original cross-classification dependence. Meanwhile, the effect is 91% for the lognormal test data. Accordingly, we see IlocA retains a sizeable degree of dependence as measured by our *hack* of the Altman Index.

The aggregated table $A^\dagger$ is not $m \times n$ and so we should like a more rigorous evaluation of IlocA's effectiveness. Specifically, by design IlocA should *only aggregate independent cells* and our second test is designed to confirm it performs accordingly. This test examines the pattern and size of the clusters formed when we employ IlocA. Indeed, if we have contingency table $A$, where the rows and columns are independent according to a $\chi^2$ test, then all the log-odds ratios will be close to zero. In this case, IlocA should combine cells into clusters based on the minimum cell size rule only. Thus, if the original table $A$ is an independence table with cell values of similar size, we should see a similar number of cells in each cluster of cells in the aggregated table $A^\dagger$. Broadly speaking, the distribution of number of cells in each cluster should be uniform under repeated applications of IlocA to independence tables. Our simulations in this section attempt to verify this hypothesis. Accordingly, if we find a uniform allocation of cells to clusters then we have demonstrated IlocA *aggregates independent cells only* and the procedure therefore controls type 1 error.

We conduct four separate simulations using the data generating processes set out above. Specifically, in each simulation we aggregate 100 independence tables, where the data are generated from the Uniform distribution and classified into either an 8-by-5 or a 10-by-8 contingency table. This process is repeated for the lognormal data generating process. We record the Cluster ID of each aggregated/coloured atomic cell on each simulated contingency table. Across all 100 simulated contingency tables we then report the average number of cells assigned a specific colour. The results obtained for the four separate simulation runs are shown in Table 1.

For contingency tables based on Uniform data the results in Table 1 show IlocA typically assigns up to 10 cluster/aggregation groups. Each cluster comprises between 2 and 3 atomic cells when the simulated contingency table is 8-by-5, while each cluster comprises between 5 and 6 atomic cells when the simulated contingency table is 10-by-8. For both table dimensions, the allocation of atomic cells to clusters is quite uniform. Moreover, the coloured table generated from the first simulation run shows no strong pattern of assignment. Clearly, since the column classification variable in each table is generated via a Uniform distribution, we should expect, (and based on the sample table we can see) roughly equal frequency counts in each atomic cell in each table. Also, each generated table is an independence table, so there is no row and column dependence. On this basis, IlocA assigns most atomic cells with a frequency count of 20 or less, to a cluster, and each cluster has roughly an equal number of similar sized atomic cells. It is clear from the results in Table 1, IlocA has not created a few clusters with many cells, but rather tends to deliver a large number of equal sized clusters in line with expectations. Importantly, this also suggest IlocA does not identify association in the table where there is none.

Table 1: Average number of cells per cluster assigned in simulated contingency tables by IlocA and sample coloured tables

| Cluster/colour ID | Cells per Aggregation group/ Cluster | | | |
|---|---|---|---|---|
| | Uniform data tables | | lognormal data tables | |
| | 8 by 5 | 10 by 8 | 8 by 5 | 10 by 8 |
| -1 | 2.3 | 4.9 | 4.7 | 13.1 |
| -2 | 2.4 | 5.0 | 4.0 | 11.9 |
| -3 | 2.5 | 5.1 | 3.8 | 8.5 |
| -4 | 2.6 | 5.3 | 3.8 | 6.7 |
| -5 | 2.6 | 5.3 | | |
| -6 | 2.6 | 5.8 | | |
| -7 | 2.7 | 5.6 | | |
| -8 | 2.7 | 5.7 | | |
| -9 | 2.6 | 5.3 | | |
| -10 | 1.6 | 4.9 | | |

Coloured table on 1st simulation run

| Uniform Prob($\chi^2$)=0.74 | | | | | | lognormal Prob($\chi$2)=0.22 | | | | |
|---|---|---|---|---|---|---|---|---|---|---|
| 10 | 12 | 12 | 13 | 13 | | 6 | 34 | 9 | 7 | 4 |
| 11 | 10 | 16 | 14 | 9 | | 0 | 30 | 16 | 11 | 3 |
| 10 | 17 | 9 | 7 | 17 | | 2 | 33 | 10 | 11 | 4 |
| 14 | 14 | 5 | 14 | 13 | | 6 | 35 | 10 | 7 | 2 |
| 15 | 11 | 13 | 15 | 6 | | 3 | 33 | 14 | 6 | 4 |
| 9 | 8 | 17 | 15 | 11 | | 5 | 33 | 13 | 5 | 4 |
| 7 | 12 | 14 | 15 | 12 | | 3 | 37 | 10 | 5 | 5 |
| 11 | 9 | 20 | 9 | 11 | | 2 | 37 | 9 | 7 | 5 |

In contrast, when we examine the results for contingency tables where the column classification variable is generated via a lognormal distribution, a different picture is evident. These tables have a single dominant column mimicking the style of a geographic table. Indeed, the sample table shows this dominance clearly in the second column. Otherwise, with a $\chi^2$ probability of 0.22 this table is an independence table. In fact, this is also true for the other 99 simulated tables, though many of these typically have a $\chi^2$ probability close to the critical region. When IlocA aggregates this style of table, as required, it leaves the dominant column alone. However, the statistics in Table 1 show IlocA typically assigns only four cluster/aggregation groups. For the 8-by-5 simulated contingency tables, each cluster comprises between 3.8 and 4.7 atomic cells. The allocation of atomic cells to clusters is therefore quite uniform, and once again, no significant pattern is evident in the sample-coloured table. Meanwhile, for simulated tables of dimension 10-by-8, each cluster comprises between 6.7 and 13.1 atomic cells. Clearly, the allocation of atomic cells to clusters is not uniform. In this instance, the second column is far less dominant than in the 8-by-5 table, simply because we are assigning the same number of observations across a table with larger dimensions. Also, since the $\chi^2$ probability for several of these tables is typically close to the critical region, some 10-by-8 tables may not really be independence tables. Accordingly, we should not expect roughly equal frequency counts in each atomic cell in each table, but rather fewer aggregated cells with a far more varied allocation of atomic cell frequency counts across those aggregated cells. On this basis, our expectation is IlocA will generate a varied number of aggregated cells, with each of these being composed of a varied number atomic cells, which themselves vary in terms of their frequency counts. Thus, where there is evidence in favour of dependence in a simulated contingency table, the non-uniform allocation of atomic cells to aggregation groups/clusters supports this expectation. Importantly, this suggest IlocA respects dependent relationships in a contingency table and aggregates/merges *independent cells only*, while leaving cells with large frequency counts or cells exhibiting dependence largely intact.

## 5. Imputation Simulation Study Framework

IlocA has been primarily formulated to create aggregation cells for imputation. Accordingly, in this section, we set out a framework for a simulation study to evaluate the performance of IlocA under eight distinct imputation settings. In each setting, 100 survey datasets are generated, each with $t = 1 - N = 480$ observations. On each dataset there is a response variable $y_t$, a response indicator variable $r_t$, with $r_t = 1$ for each respondent and $r_t = 0$ for a non-response. Depending on the imputation setting, there are either two auxiliary classification variables $z_{1t}$ and $z_{2t}$, or three continuous auxiliary variables $z_{1t}$, $z_{2t}$ and $z_{3t}$. In the latter case, we use the first two of these variables to define the pair of auxiliary classification variables $z_{1c,t}$ and $z_{2c,t}$, while the third variable is used to define one of the response models. Without loss of generality, we assume survey weights are all equal to 1, so we have a census. The pair of auxiliary classification variables are cross-classified to give the contingency table of observation counts. IlocA is applied to this contingency table to create a set of aggregated cell clusters – these clusters are our imputation cells. Cell mean estimates are computed for the responses (*i.e.*, observations where $r_t = 1$) within these imputation cells. The within imputation cell mean is then applied to each missing response falling within that imputation cell. Clearly, the within imputation cell estimate of the mean $\bar{y}_{I,c}$ is given by

$$\bar{y}_{I,c} = \bar{y}_{c_r}$$

where $\bar{y}_{c_r}$ is the mean of the response values within imputation cell $c$. We define the imputed estimator $\bar{Y}_I$, of the overall dataset mean $\bar{Y}$ of the response variable $y_t$, based on $C$ imputation cells to be

$$\bar{Y}_I = \sum_{c=1}^{C} \frac{N_c}{N} \bar{y}_{I,c} \qquad (5)$$

where $N_c$ is the observation count in cell $c$. We note, Haziza & Beaumont (2007), show (a weighted version of) this cell mean estimator is unbiased and its variance is small, when the auxiliary variables explain the variability of the response variable $y_t$.

Two separate data generating processes are used to create variable values on each imputation setting survey dataset.

Data Generating Process 1 is a purely random model. The pair of auxiliary classification variables $z_{1t}$ and $z_{2t}$ are generated from uniform data according to the procedure outlined in the previous section based on Kass (1980). These variables are cross-classified into an $8 \times 5$ contingency table of atomic cells. IlocA is applied to these atomic cells with a count below twenty. Response variable values $y_t$ are generated according to:

$$y_t = u_t^{5/4} \qquad (6)$$

where $u_t$ is a standard uniform variate value. Experimentally, we have chosen this form for $y_t$, so that when $y_t$ is regressed against the constant indictor variable, we get a model $R^2$ (coefficient of determination) of about 0.67 – the model explains 2/3$^{\text{rds}}$ of the variation in $y_t$.

Data Generating Process 2 is a linear model given in Haziza, & Beaumont (2007). Here, the response data are generated according to

$$y_t = \beta_0 + \beta_1 z_{1t} + \beta_2 z_{2t} + \beta_3 z_{1t}^2 + \varepsilon_t \qquad (7)$$

so, the response variable value is a (random) function of the auxiliary variable values. Continuous auxiliary variables $z_{1t}$, $z_{2t}$ and $z_{3t}$, are each generated from an exponential distribution with mean 100. We set the parameters in (7) to their values given in Haziza, & Beaumont (2007), specifically $\beta_0 = 20$, $\beta_1 = 10$, $\beta_2 = 0.5$ and $\beta_3 = 10$. The random noise $\varepsilon_t \sim N(0, 200)$, is chosen experimentally, so that when $y_t$ is regressed against the constant indictor variable, $z_{1t}$, $z_{2t}$ and $z_{1t}^2$, we once again get a model $R^2$ (coefficient of determination) of about 0.67. Meanwhile, to obtain the pair of auxiliary classification variables, we first create a temporary auxiliary variable $z_{1t}^* = z_{1t} + z_{1t}^2$. We take this variable and the continuous auxiliary variable $z_{2t}$ and separately group each of these variables into five classes. The classes are determined based on the cumulative discrete distribution [0.00, 0.04, 0.10, 0.15, 0.20, 1.0] – this distribution was found experimentally and ensures all five classes are not empty. Following along similar lines to Haziza & Beaumont (2007), the resulting pair of auxiliary classification variables $z_{1c,t}$ and $z_{2c,t}$ are cross-classified into a $5 \times 5$ contingency table of atomic cells. IlocA is applied to these atomic cells with a count below twenty. We note, utilising this temporary variable ensures the data model based on $z_{1c,t}$ and $z_{2c,t}$ is correctly specified and moreover meaningfully explains the response variable.

We also generated a non-response to variable $y_t$ according to two response models given in Haziza, & Beaumont (2007).

<u>Response Model 1</u> defines the response probability $p_t$, for observation $t$, according to the logistic function

$$\log\left(\frac{p_t}{1-p_t}\right) = \lambda_0 + \lambda_1 z_{1,t} + \begin{cases} \lambda_2 z_{2t} & \text{Data Generating Process 1} \\ \lambda_2 z_{3t} & \text{Data Generating Process 2} \end{cases} \quad (7)$$

Under both model forms, the probability of response is a function of the auxiliary variables only, so the response mechanism is ignorable.

<u>Response Model 2</u> defines the response probability $p_t$, for observation $t$, according to the logistic function

$$\log\left(\frac{p_t}{1-p_t}\right) = \lambda_0 + \lambda_1 y_t + \lambda_2 z_{3t} \quad (8)$$

Importantly, under this model the probability of response is a function of the auxiliary variable $z_{3,t}$ and the response itself, so this response mechanism is therefore non-ignorable. Under both response models, we then generate response indicator values $r_t = 0, 1$, as Bernoulli r.v.'s using the respective response probabilities $p_t$.

The combination of two data generating process and two response models defined above, gives rise to four possible imputation settings. Finally, we further expand these four settings to eight, by choosing response model parameter values to give mean response rates of 75% and 50% respectively. We evaluate the imputation performance of IlocA under each of the eight imputation settings – the response model parameter values chosen for each of the eight imputation settings are shown in Table 2.

Table 2: Response Model Parameter settings

| | Response Model Parameters | | | |
|---|---|---|---|---|
| Data Generating Process 1 (Uniform Model) | | | | |
| | Response Rate (%) | Parameters | | |
| | | $\lambda_0$ | $\lambda_1$ | $\lambda_2$ |
| Response Model 1 | 75 | 0.05 | 0.22 | 0.05 |
| | 50 | 0.05 | -0.05 | 0.05 |
| Response Model 2 | 75 | 0.05 | 1.00 | 0.30 |
| | 50 | 0.05 | 1.00 | -0.15 |
| Data Generating Process 2 (Regression Model - Haziza, & Beaumont, 2007) | | | | |
| | Response Rate (%) | Parameters | | |
| | | $\lambda_0$ | $\lambda_1$ | $\lambda_2$ |
| Response Model 1 | 75 | 1.00 | 0.50 | 0.10 |
| | 50 | 1.00 | -12.00 | -100.00 |
| Response Model 2 | 75 | 0.10 | 0.04 | 0.05 |
| | 50 | 0.10 | 0.08 | -200.00 |

In each imputation settings, the bias and root mean square error of the overall imputed mean estimate $\bar{Y}_I$, of the overall dataset mean $\bar{Y}$, is measured. Specifically, in each setting we have 100 simulation datasets, each with 480 observations. On each dataset we add a working response variable $w_t = y_t$ when $r_t = 1$, and $w_t =$ missing when $r_t = 0$. Within imputation cells, we impute a value for each missing $w_t$ using the cell mean imputation method. This gives us a full set of 480 actual and imputed $w_t$ values. We compute the imputed estimate of the mean $\bar{Y}_I$ using this full set. In parallel, we compute actual overall dataset mean $\bar{Y}$ using the full set of response values $y_t$ on that simulated dataset. As a measure of the bias of $\bar{Y}_I$, we compute the Monte Carlo percent relative bias (RB) over all 100 simulated datasets, this quantity is given by

$$\text{RB}(\bar{Y}_I) = \frac{E_{MC}(\bar{Y}_I) - \bar{Y}^{(k)}}{\bar{Y}} \times 100\% \tag{8}$$

where $E_{MC}(\bar{Y}_I) = \frac{1}{100}\sum_{k=1}^{100} \bar{Y}_I^{(k)}$, and $\bar{Y}_I^{(k)}$ and $\bar{Y}^{(k)}$ denote the imputed estimator and actual overall dataset mean respectively, from the $k^{th}$ simulation dataset. Also, as a measure of the variability of $\bar{Y}_I$, we compute the Monte Carlo percent relative root mean square error (RRMSE) given by

$$\text{RRMSE}(\bar{Y}_I) = \frac{\sqrt{MSE_{MC}(\bar{Y}_I)}}{\bar{Y}^{(k)}} \times 100\% \tag{9}$$

where $MSE_{MC}(\bar{Y}_I) = \frac{1}{100}\sum_{k=1}^{100}\left(\bar{Y}_I^{(k)} - \bar{Y}^{(k)}\right)^2$.

5.1 Bias monitoring

Deterministic regression measure

We implement a deterministic regression approach to gauge the relative efficiency of IlocA, Null and Atomic cell based imputed overall mean estimates $\bar{Y}_I$. Specifically, for Data Generating Process 1, we fit the deterministic regression model form

$$y_t = \beta_0 + \beta_1\left(\frac{t}{480}\right) \tag{8}$$

to the ordered values of $y_t$ when $r_t = 1$. Meanwhile, for Data Generating Process 2, we fit the functional form given in equation (7) to the response values. Once again, we create a working response variable $w_t = y_t$ when $r_t = 1$, and when $r_t = 0$ we predict $w_t$ using the estimated deterministic regression model. We use $w_t$ to arrive at a baseline mean estimate $\bar{Y}_I$ and its associated RB. When the postulated deterministic regression model holds, Haziza, & Beaumont (2007) say this RB value provides a baseline to assess the quality of the RB produced by other cell-based imputation procedures.

Response quantile imputation cells measures

Clearly, on each simulation dataset the full 480 response values $y_t$ and response indicators $r_t$ are available. Accordingly, we can use the full set of response values to create actual response variable quantiles for $y_t$. We construct quantiles using all 480 values of $y_t$ and employ them to define response quantile imputation cells. Once again, we generate an overall mean estimate $\bar{Y}_I$ based on working response values $w_t$ falling into these response quantile imputation cells. In each imputation setting and on each simulated dataset, we consider seven different sets of quantiles that range from five quantiles to thirty-five quantiles in steps of five. Since response variable quantiles for the full 480 response values $y_t$ are exact quantiles, this provides a second framework to monitor the bias of the imputed estimator. In general, this bias will decrease as the number of response quantile imputation cells increases. Accordingly, the minimum number of response quantile cells required to achieve a stable overall mean estimate $\bar{Y}_I$, provides an estimate of the optimal number of cells required to impute missing responses values.

5.2 Sources of error

Of course, in practice, imputation cells are formed after data are collected and the survey dataset formed. Accordingly, when we apply a specific method to determine the set of imputation cells to this survey dataset, the resulting cell boundaries and the number of imputation cells will be fixed. However, we are interested in evaluating the imputation cell generating mechanism of IlocA, and so we introduce uncertainty into the cell generating mechanism by randomly varying the survey dataset variables on each simulated survey dataset. Thus, when we employ the resulting set of imputation cells to impute missing values, we generate imputation error associated with the imputation cell generating mechanism itself.

In contrast, it is important to emphasise, when we employ the cell mean imputation method there is no random component associated with this imputation method, as would be the case if we employed a *hot-deck* method within each imputation cell. This means there is no imputation error associated with the cell mean imputation method itself. Our evaluation of IlocA is therefore pure, as imputation error due to the cell generating mechanism, is not conflated with imputation error due to the imputation method employed. For analysis purposes, this is a key benefit of choosing to work with the cell mean imputation method.

## 6. Imputation Simulation Study Results and Discussion

6.1 Results under correct data generating process specification

Tables 3 and 4 show the RB and RRMSE results we obtained under each of the eight separate simulation settings. We give results for cell mean imputation where we employ either imputation cells generated by IlocA, or we employ the fully aggregated single Null cell, or we utilise the detailed set of Atomic cells. To frame our discussion and evaluate the relative efficiency of imputation cells generated by IlocA, we also give bias monitoring results obtained when the deterministic regression method is employed, and when exact response quantile cells are employed.

First, looking at the baseline exact measure results in Table 3 for deterministic regression, we see the RB is close to zero while the RRMSE is less than 0.12% in all four imputation settings. Furthermore, examining the full response quantile cells, both the RB and RRMSE stabilise at twenty-five cells across both response models and both response rate settings. At twenty-five cells, the RB and RRMSE stabilise at maximum values of 0.03% and 0.23% respectively, values close to those obtained with deterministic regression. This shows the likely minimum number of cells required to satisfactorily impute these uniformly distributed missing data is twenty-five cells. We note, this is not consistent with the suggestion made by Eltinge & Yansaneh, (1997) and Rosenbaum & Rubin, (1983), that five cells may provide the most effective bias reduction. Nonetheless, this in line with the observation of Haziza, & Beaumont (2007), "that a moderate number of cells ($C < 25$) provides an effective bias reduction, which in turn may lead to relatively stable estimates since the number of responses in each cell is likely to be fairly large". Counterintuitively, RB values obtained when five full response quantile cells are employed are relatively small. However, this level of bias reduction does not seem to be robust to moderate changes in number of cells, when the number of cells chosen is relatively small. This inconsistency suggests employing only five or ten cells may be unwise in practice.

In Table 3, the RB results for IlocA under each of the four response model and response rate settings range from 0.07% to a maximum of 1.71%. These figures compare favourably with their exact measure counterparts (deterministic regression and response quantile cells). We also measured the average number of cells generated by IlocA across the 100 simulated datasets. This turned out to be 26½ cells, a value close to the optimal twenty-five. Thus, it would seem twenty-five cells are required to accurately impute these missing data, while simultaneously maintaining any dependence that exists between the auxiliary classification variables and the response on the dataset. Turning to the RB generated when we employ Atomic cells, the maximum RB is 0.36%. In this instance there are 40 atomic cells, so it is not surprising the RB is close to optimal. However, at the other extreme, employing a single Null cell gives RB values ranging from 1.1% to 9.7%. As might be expected, employing a Null cell, or equivalently the overall response mean value, gives poor results. Of course, this naïve approach to imputation is rarely, if ever, used in practice.

Table 3: Imputation results based on data generated from a Uniform distribution

| Cell Structure | Response Model 1 | | | Response Model 2 | |
|---|---|---|---|---|---|
| | RB (%) | RRMSE (%) | | RB (%) | RRMSE (%) |
| **Response Rate 75%** | | | | | |
| **IlocA** | **0.07** | **0.84** | | **1.71** | **1.97** |
| Null | 1.10 | 1.90 | | 9.56 | 9.70 |
| Atomic | -0.01 | 0.37 | | 0.14 | 0.35 |
| *Exact Measures* | | | | | |
| *Deterministic Regression* | *-0.01* | *0.06* | | *-0.01* | *0.11* |
| *Full Response Quantile Cells* | | | | | |
| *No. of cells* | | | | | |
| *5* | *-0.60* | *1.63* | | *6.93* | *7.07* |
| *10* | *-4.86* | *5.02* | | *0.31* | *0.99* |
| *15* | *-7.63* | *7.71* | | *-3.71* | *3.16* |
| *20* | *-4.10* | *4.21* | | *-1.40* | *1.51* |
| *25* | *0.01* | *0.08* | | *0.03* | *0.09* |
| *30* | *-0.03* | *0.16* | | *0.03* | *0.12* |
| *35* | *0.02* | *0.12* | | *0.05* | *0.13* |
| **Response Rate 50%** | | | | | |
| **IlocA** | **0.45** | **1.47** | | **0.88** | **1.63** |
| Null | 2.54 | 3.98 | | 3.11 | 4.21 |
| Atomic | -0.08 | 0.77 | | 0.36 | 0.74 |
| *Exact Measures* | | | | | |
| *Deterministic Regression* | *0.00* | *0.10* | | *-0.01* | *0.12* |
| *Full Response Quantile Cells* | | | | | |
| *No. of cells* | | | | | |
| *5* | *-2.10* | *3.62* | | *-1.35* | *3.03* |
| *10* | *-13.55* | *13.77* | | *-12.23* | *12.43* |
| *15* | *-18.79* | *18.90* | | *-17.51* | *17.61* |
| *20* | *-8.88* | *8.97* | | *-8.29* | *8.38* |
| *25* | *0.01* | *0.23* | | *0.03* | *0.15* |
| *30* | *-0.04* | *0.23* | | *-0.03* | *0.22* |
| *35* | *-0.01* | *0.22* | | *0.00* | *0.22* |

Looking the RRMSE, the patterns observed in relation to the RB are repeated. That is, employing Atomic cells gives RRMSE values close to optimal deterministic regression and full response quantile values. IlocA, meanwhile gives estimated overall mean values that have a slightly higher level of variation. More generally, under response model 1, where the missing data mechanism is ignorable, the RB value for IlocA changes from 0.07% to 0.45%, a 0.38% increase, when the response rate decreases from 75% to 50%. Correspondingly, the RRMSE values increase by 0.63%. Meanwhile, under response model 2, where the missing data mechanism is non-ignorable, both the RB and RRMSE fall slightly. This does not happen when Atomic cells are employed, suggesting IlocA is more robust to changes in the response model specification.

Table 4 gives the results obtained when data are generated according to the regression model given of Haziza & Beaumont (2007). Examining the baseline exact measure results for deterministic regression, we see the RB is less than 0.56% in absolute terms under response model 1. The RRMSE under this response model is somewhat larger, but below 3.2%. However, under response model 2 where the missing data mechanism is non-ignorable, the RB is 10.83% when the response level is 75% and 24.48% when the response level is 50%. RRMSE values meanwhile are a similar size to RB values. Despite the fact these values seem large, they too are in line with the results obtained by Haziza & Beaumont (2007) for this imputation setting. Indeed, they comment "the behaviour of the RRMSE is very close to the behaviour of the RB because of the presence of substantial bias".

Examining the full response quantile cell results in Table 4, once again both the RB and RRMSE stabilise at twenty-five cells across both response models and both response rate settings. At twenty-five cells, the RB and RRMSE stabilise at maximum values of 0.03% and 0.50% respectively under response model 1. While under response model 2 the corresponding values are 0.73% and 1.80%.

Table 4: Imputation results based on data generated from a Haziza's Regression model

| Cell Structure | Response Model 1 | | | Response Model 2 | |
|---|---|---|---|---|---|
| | RB (%) | RRMSE (%) | | RB (%) | RRMSE (%) |
| **Response Rate 75%** | | | | | |
| **IlocA** | **-0.26** | **1.91** | | **10.82** | **11.06** |
| Null | -0.28 | 1.87 | | 10.72 | 10.98 |
| Atomic | -0.23 | 1.98 | | 9.20 | 9.38 |
| *Exact Measures* | | | | | |
| *Deterministic Regression* | *-0.25* | *1.89* | | *10.73* | *11.00* |
| *Full Response Quantile Cells* | | | | | |
| *No. of cells* | | | | | |
| *5* | *-2.83* | *3.30* | | *5.54* | *5.96* |
| *10* | *-6.86* | *7.02* | | *-2.41* | *2.77* |
| *15* | *-8.78* | *8.87* | | *-5.29* | *5.45* |
| *20* | *-4.96* | *5.06* | | *-2.44* | *2.62* |
| *25* | *-0.02* | *0.25* | | *0.18* | *0.39* |
| *30* | *-0.14* | *0.34* | | *0.20* | *0.44* |
| *35* | *0.04* | *0.28* | | *0.36* | *0.53* |
| **Response Rate 50%** | | | | | |
| **IlocA** | **-0.67** | **3.27** | | **24.51** | **24.73** |
| Null | -0.54 | 3.05 | | 24.63 | 24.81 |
| Atomic | -0.60 | 3.24 | | 24.44 | 24.66 |
| *Exact Measures* | | | | | |
| *Deterministic Regression* | *-0.56* | *3.04* | | *24.62* | *24.80* |
| *Full Response Quantile Cells* | | | | | |
| *No. of cells* | | | | | |
| *5* | *-7.17* | *7.74* | | *18.51* | *18.99* |
| *10* | *-17.29* | *17.44* | | *-1.41* | *4.11* |
| *15* | *-19.67* | *19.77* | | *-8.90* | *9.33* |
| *20* | *-9.88* | *9.97* | | *-3.95* | *4.28* |
| *25* | *-0.04* | *0.50* | | *0.73* | *1.80* |
| *30* | *-0.27* | *0.55* | | *0.57* | *1.46* |
| *35* | *0.01* | *0.50* | | *0.78* | *1.56* |

In Table 4, the RB results for IlocA under response model 1 are -0.24% and -0.57%, when the response rates are 75% and 50% respectively. The corresponding RRMSE values are 1.97% and 3.20%. Similar RB and RRMSE results are replicated when Atomic cells and deterministic regression is employed. Intriguingly, using a single Null cell also gives RB and RRMSE that are close to IlocA. When we measured the average number of cells generated by IlocA across the 100 simulated datasets, this came out at 19 cells, a value somewhat below the optimal twenty-five. Accordingly, it seems the number of cells employed to impute these missing data has little effect on the bias or variance, whether it be a single Null cell, the 19 cells generated by IlocA or indeed the set of 40 atomic cells. Nevertheless, it is comforting to see both the RB and RRMSE are relatively small when the response model is ignorable. When we examine what happens under non-ignorable response model 2, the pattern evident for response model 1 is replayed, except the RB and RRMSE values are considerably larger. Clearly, these results show the quality of the imputed estimator is determined solely by the substantial bias associated with accurately modelling Data Generating Process 2. Consequently, RB and RRMSE values are relatively large irrespective of the imputation approach employed – a finding echoed in Haziza & Beaumont (2007).

6.2 Results under model mis-specification

In Table 5 we extend our analysis to see what, if any, is the impact when the data generating process and associated cell structure is mis-specified. For example, atomic cells created under Data Generating Process 1, arise from a pair of auxiliary classification variables $z_{1t}$ and $z_{2t}$, where $z_{2t}$ is generated from uniform data. The response variable $y_t$ is also a function of this uniform random variate, so, in particular, $z_{2t}$ explains the response. We mis-specify $z_{2t}$ by generating it from a lognormal distribution as described in section 4, specifically, $z_{2t} = int(mod(1 + v_t, n))$ and $v_t \sim$ lognormal(0,2). As before, we then cross-classify $z_{1t}$ and $z_{2t}$ into an 8 × 5 contingency table of atomic cells. For the Data Generating Process 2, we mis-specify the classification variable $z_{1c,t}$ by excluding the quantity $z_{1t}^2$ from the definition of the temporary variable $z_{1t}^*$ used to create $z_{1c,t}$. The results obtained from running IlocA under these mis-specified models is given in Table 5.

Table 5: Imputation results under mis-specified models

| Cell Structure | Uniform Data Generation Process $z_2$ specified with lognormal | | | | Regression Model Data Generation Process $z_1^2$ excluded from $z_{1c}$ | | | |
|---|---|---|---|---|---|---|---|---|
| | Response Model 1 | | Response Model 2 | | Response Model 1 | | Response Model 2 | |
| | RB (%) | RRMSE (%) | RB (%) | RRMSE (%) | RB (%) | RRMSE (%) | RB (%) | RRMSE (%) |
| **Response Rate 75%** | | | | | | | | |
| **IlocA** | 0.99 | 2.08 | 9.19 | 9.36 | -0.24 | 1.97 | 10.77 | 11.03 |
| *Deterministic Regression* | *0.01* | *0.07* | *0.03* | *0.11* | *-0.25* | *1.89* | *10.83* | *11.08* |
| **Response Rate 50%** | | | | | | | | |
| **IlocA** | 2.17 | 3.80 | 2.63 | 4.02 | -0.57 | 3.20 | 24.48 | 24.68 |
| *Deterministic Regression* | *0.00* | *0.12* | *0.01* | *0.12* | *-0.56* | *3.04* | *24.62* | *24.80* |

When we compare the results for the Data Generation Process 1 under response model 1 in Table 5, to those given in Table 3, we see the mis-specified model RB and RRMSE figures in Table 5 are slightly larger, but still remain relatively small. IlocA does not seem to be sensitive to model mis-specification in this instance. However, as we move to response model 2, where the response mechanism is non-ignorable, results are less consistent as we get larger RB and RRMS values at the higher response rate of 75%. Nonetheless, the levels of RB and RRMSE remain moderate, suggesting IlocA is not too sensitive to model misspecification in this instance either.

Turning to the results under the regression model Data Generation Process 2, we see a somewhat different picture. Under response model 1, the RB and RRMSE values for IlocA are almost identical to corresponding figures in Table 4. Accordingly, how well the cells generated by IlocA explain the response has little impact on the quality of the estimated overall mean. Of course, RB and RRMSE levels remain relatively low, showing the cells generated by IlocA provide a satisfactory explanation of the response, when the non-response mechanism is ignorable. Meanwhile, when we compare the results under response model 2, we see the pattern evident for response model 1 is repeated, albeit with much higher RB and RRMSE values. As we noted above, these large error levels are directly due to the substantial bias. That said, IlocA performs consistently with respect to response levels. Specifically, the RB and RRMSE values come in at about 10% when the response rate in 75%, and then rise to about 25% when the response rate falls to 50% - this is comforting. In any event, the results obtained for IlocA in this setting compare favourably with those obtained using deterministic regression, and indeed, those obtained by Haziza & Beaumont (2007), regardless of whether imputation cells are specified correctly or not.

6.3 Summary remarks concerning our Imputation Simulation Results

We highlight the key outcomes from our imputation simulation:

Our analysis based on exact response quantiles has demonstrated that twenty-five cells are required reduce the bias to a maximum extent. With 480 observations and a 50% response rate, this indicates a minimum of about 10 responses per cell are needed to impute missing data satisfactorily. This seems to be true whether missingness is ignorable or not.

Generally, it seems clear that imputation based on five or ten cells is unwise, as the level of bias reduction is not robust to small changes in the number of cells.

When the bias is substantial, the cell-based imputation and deterministic regression approaches achieved no efficiency gains.

IlocA generated an aggregated cell structure with between 19 and 26 cells depending on the imputation setting. With this structure, it gave RB and RRMSE values similar to deterministic regression and exact response quantile cells. This level of performance is achieved while ensuring dependencies between the classification variables and the response is maintained.

IlocA gives consistent results since a decrease in the response level generally gives higher RB and RRMSE values.

The performance of IlocA compares favourably with other approaches regardless of whether the imputation cells are specified correctly or not. Accordingly, IlocA seems to be robust to imputation cell misspecification.

## 7. Case Study

In this section we give a flavour of the performance of IlocA in an applied setting. Specifically, we focus on estimating energy ratings of detached residential dwellings in county Wicklow in Ireland. Policy interest focusses on this subset of dwellings, as many of them are in rural areas and are older dwellings. Accordingly, they are likely to have poor energy performance and require substantial investment to retrofit to achieve better levels of energy efficiency. We use the Energy Performance Certificate (EPC) measurement, known in Ireland as the Building Energy Rating (BER) of each dwelling. The compendium of BER measurements is compiled by the Sustainable Energy Authority of Ireland into an administrative dataset called the BER dataset. The key variable of interest on this dataset is the Energy Value (EV) of each dwelling expressed in kW hours/m$^2$/year. This value is arrived at by a trained assessor who measures attributes of the building and inputs these into a standard procedure that computes the EV. This procedure is called the DEAP (Dwelling Energy Assessment Procedure) and is overseen by the Sustainable Energy Authority of Ireland. While many attributes are measured, a few that are common to all buildings, such as dwelling type (house, flat etc.), type of space heating and age of dwelling are critical inputs to determine the EV.

Separately, work is ongoing to match the BER and Census 2016 datasets by Census household reference number. For detached residential dwellings in county Wicklow, about 27%, out of a total of just over 17,143 households have been matched. Accordingly, we seek to impute the EV for the remaining 73% of households. Of course, given this level of missingness, alternative strategies such as statistical matching (see D'Orazio et. al., 2006) or re-weighting (see Curtis et. al. 2015) may be employed. That said, we proceed and report our findings based on applying cell mean imputation with cells generated by IlocA.

Clearly, the Census dataset has many variables that may be used to arrive at a model to impute missing EVs. However, the majority of these are irrelevant and/or useless to predict EVs. According to the DEAP, the reason for this is the EV measurement itself deliberately excludes personal household energy consumption for lighting, cooking, washing etc. Only building related energy consumption, such as space heating or floor area are included. This means Census variables, such as the number of persons in a household, or their ages, or their occupations are neither measured nor particularly relevant. Only dwelling variables, such as type of heating, year of construction and geographic location are therefore relevant. On this basis, we proceed to identify a set of imputation cells founded on type of heating, year of construction and location of each dwelling. First, we collapse heating type into two classes (oil and other heating) and year of construction into four classes (before 1960, 1961-1979, 1980-1999, 2000 or later). These are then combined to give a new variable with eight classes. Second, we create an aggregated geographic classification variable based on the number of households in each of the 84 electoral divisions in county Wicklow. We created a new classification variable 'Electoral Division Size', by mapping individual electoral division codes to ten electoral division size classes, based on the number of households in each electoral division. The resulting classification has three large size classes, two with over 5,000 households and a third with about 2,700 households, covering the main towns located on the east coast. Six of the remaining classes had between 300 and 600 households relating to rural areas. Table 6 (top) shows the frequency count of all Census households cross-classified by the pair of variables

Table 6: Electoral Division Size by Year of Construction by Heating Type Contingency Table

| | All households | | | | | | | | |
|---|---|---|---|---|---|---|---|---|---|
| | Year of Construction | | | | | | | | |
| | Before 1960 | | 1960-1979 | | 1980-1999 | | 2000 or after | | |
| | Heating Type | | | | | | | | |
| Electoral Division Size | Oil | Other | Oil | Other | Oil | Other | Oil | Other | Total |
| 1 | 354 | 705 | 476 | 541 | 856 | 533 | 272 | 1,473 | 5,210 |
| 2 | 222 | 210 | 439 | 714 | 495 | 1,181 | 119 | 1,893 | 5,273 |
| 3 | 280 | 91 | 293 | 129 | 510 | 389 | 370 | 627 | 2,689 |
| 4 | 155 | 45 | 109 | 17 | 209 | 94 | 469 | 394 | 1,492 |
| 5 | 60 | 22 | 53 | 31 | 109 | 19 | 67 | 38 | 399 |
| 6 | 45 | 17 | 37 | 11 | 83 | 19 | 88 | 19 | 319 |
| 7 | 83 | 28 | 42 | 8 | 97 | 28 | 227 | 65 | 578 |
| 8 | 54 | 16 | 30 | 11 | 76 | 13 | 120 | 47 | 367 |
| 9 | 41 | 12 | 9 | 8 | 68 | 16 | 129 | 46 | 329 |
| 10 | 87 | 29 | 25 | 9 | 92 | 25 | 148 | 72 | 487 |
| Total | 1,381 | 1,175 | 1,513 | 1,479 | 2,595 | 2,317 | 2,009 | 4,674 | 17,143 |
| | Matched households | | | | | | | | |
| | Year of Construction | | | | | | | | |
| | Before 1960 | | 1960-1979 | | 1980-1999 | | 2000 or after | | |
| | Heating Type | | | | | | | | |
| Electoral Division Size | Oil | Other | Oil | Other | Oil | Other | Oil | Other | Total |
| 1 | 86 | 223 | 141 | 184 | 258 | 183 | 75 | 417 | 1,567 |
| 2 | 61 | 73 | 116 | 240 | 144 | 405 | 27 | 655 | 1,721 |
| 3 | 58 | 14 | 69 | 28 | 125 | 94 | 66 | 175 | 629 |
| 4 | 17 | 6 | 20 | 5 | 54 | 16 | 104 | 89 | 311 |
| 5 | 3 | 4 | 7 | 31 | 4 | 9 | 10 | - | 68 |
| 6 | 7 | 4 | 6 | 3 | 11 | 19 | 3 | - | 53 |
| 7 | 5 | 3 | 9 | 1 | 18 | 2 | 29 | 7 | 74 |
| 8 | 11 | 4 | 7 | 11 | 2 | 8 | 2 | - | 45 |
| 9 | 4 | 2 | 2 | 2 | 6 | 1 | 14 | 5 | 36 |
| 10 | 12 | 4 | 3 | 1 | 10 | 3 | 15 | 4 | 52 |
| Total | 264 | 337 | 380 | 506 | 632 | 740 | 345 | 1,352 | 4,556 |

Electoral Division Size and Year of Construction by Heating Type – this cross-classification generates 80 atomic cells. The geographic nature of this table is evident with urban areas in size classes 1 – 3, while rural electoral divisions are in size classes 4 – 10. The concentration of households is in the top right-hand corner of the table, that is, within urban areas where dwellings are constructed since 1980. Meanwhile, the bottom part of Table 6 shows the frequency count of matched households on the Census and BER datasets, falling into this set of 80 atomic cells. We apply IlocA to the matched cases with a view to imputing the missing EV values for unmatched cases.

Table 7 shows the results obtained when IlocA is applied to the matched households contingency table. As expected, larger cells relating to urban areas are not merged to form an aggregated cell. Thus, IlocA has respected the general urban rural geographic nature of the table. This is important in this context as detached dwellings in rural areas tend to be less energy efficient. Aggregation of smaller cells, however, is substantial – these aggregated cells are coloured and have a negative value inserted to identify the aggregated cell group. Groups labelled -1, -2, -3, -6 and -11 form well defined larger clusters of cells, generally based on row or column groups of cells. However, these groups are not simple joins of row or columns that may be associated with a one-way fixed effects model. Rather, these groups are partial combinations of cells across rows and columns that respect tabular dependence. This, being achieved while making sure each aggregated cell, has a sufficient (observation) count to ensure imputed values based on these aggregated cells are robust. In all, IlocA generates a cell structure with 35 imputation cells, comprising original large cells and aggregated cells. Given our earlier simulation study results, this number of cells seems a good compromise to ensure sufficient bias reduction while controlling variance associated with employing this cell structure.

Table 7: IlocA aggregated coloured cells for matched households

| | Matched households | | | | | | | |
|---|---|---|---|---|---|---|---|---|
| | Year of Construction | | | | | | | |
| | Before 1960 | | 1960-1979 | | 1980-1999 | | 2000 or after | |
| | Heating Type | | | | | | | |
| Electoral Division Size | Oil | Other | Oil | Other | Oil | Other | Oil | Other |
| 1 | -9 | 223 | 141 | 184 | 258 | 183 | 75 | 417 |
| 2 | 61 | -8 | -14 | 240 | -12 | 405 | -12 | 655 |
| 3 | -15 | -12 | -10 | 184 | -4 | 94 | -7 | 175 |
| 4 | -6 | -2 | 20 | 240 | -5 | -11 | 104 | -11 |
| 5 | -3 | -8 | -14 | -6 | 31 | -11 | -4 | -13 |
| 6 | -6 | -2 | -5 | -14 | -9 | -11 | 19 | -11 |
| 7 | -6 | -2 | -8 | -9 | 18 | -11 | -1 | -1 |
| 8 | -3 | -2 | -8 | -9 | -11 | -3 | -1 | -7 |
| 9 | -3 | -3 | -3 | -14 | -11 | -12 | -13 | -1 |
| 10 | -3 | -2 | -3 | -1 | -11 | -3 | -13 | -11 |

Curtis et. al (2015) used a two-way main effects regression model based on year of construction, type of heating and dwelling type, to adjust BER estimates to 2011 Census marginal totals. With a view to providing a baseline against which to compare estimates, we have applied their approach to impute missing energy values on our Census 2016 dataset. In Table 8, we classify theses EVs into seven standard energy rating bands, labelled A – G. Table 8 also gives the estimates we obtained when missing energy values are imputed using cell mean imputation within imputation cells generated by IlocA. Here, mean energy values from matched households within each of these imputation cells, are applied to households with missing energy values in the same imputation cell. Specifically, we compute a working EV equal to the original EV value for non-missing EVs, while the working value is set equal to the imputed cell mean value for missing EVs. Once again, we classify theses working EVs into the seven standard energy rating bands. In Table 8 we show the percent relative error between estimates obtained under IlocA's imputation cells and the baseline estimates of Curtis et. al.

(2015). Tables 8 also gives relative error estimates obtained when we employ the full set of 80 atomic cells.

Table 8: Estimated Building Energy Ratings for Detached Dwelling in county Wicklow

|  |  | Curtis | | % Relaative Error | | | |
|---|---|---|---|---|---|---|---|
|  |  |  |  | IlocA | | Atomic cell | |
| Energy Rating |  | No of households | Mean Energy Value | No of households | Mean Energy Value | No of households | Mean Energy Value |
| A | less than 75 | 97 | 59 | 0.0 | 0.0 | 24.7 | 0.7 |
| B | 75 - <125 | 1,491 | 120 | -9.3 | 9.2 | 4.4 | 3.2 |
| C1 | 150-<175 | 4,901 | 160 | -20.8 | 0.4 | -2.8 | 4.2 |
| C2 | 175 - <200 | 3,048 | 191 | -1.5 | -1.7 | -34.5 | 1.4 |
| C3 | 200 - <225 | 1,961 | 216 | 43.7 | -0.6 | -14.5 | -2.6 |
| D1 | 225 - <260 | 2,613 | 237 | -28.1 | 2.5 | 5.4 | 1.3 |
| D2 | 260 - <300 | 2,279 | 277 | -1.6 | -0.4 | 9.6 | -0.4 |
| E1 | 300 - <340 | 324 | 320 | 345.7 | -0.3 | 258.0 | -0.1 |
| E2 | 340 - <380 | 155 | 362 | 0.0 | 0.0 | 44.5 | -0.1 |
| F | 380 - <450 | 130 | 413 | 0.0 | 0.0 | 66.9 | 1.4 |
| G | 450 or higher | 144 | 610 | 0.0 | 0.0 | 22.2 | -2.8 |
| All |  | 17,143 | 206 | 0.0 | 5.2 | 0.0 | 5.9 |

Examining the estimates in Table 8, it is clear there is a considerable level of agreement across the three imputation approaches in relation to estimated mean energy values, as 19 of the 22 errors are less than 3% in absolute terms. However, when we examine the errors relating to the number of households, a very different picture is evident. For rating category E1, IlocA has an error of 345.7%. Thus, if estimates from IlocA are correct, there are likely far more households living in E1 rated detached dwellings in county Wicklow, than might have been heretofore suspected. This effect is balanced by 20.8% fewer households living in C1 rated dwellings. Meanwhile, results obtained when Atomic cells are employed bear our findings in relation to category E1. When combined, categories C2, C3 and D1 are broadly in balance when IlocA estimates of the number of households are compared to Curtis. This suggests a two-way main effect is likely to apply across the marginal for these categories. Interestingly, IlocA estimates also agree with those obtained under Curtis et. al's two-way model, in the tail categories E2 –

G of the energy value distribution. This too indicates a two-way main effect applies in the upper tail of the energy value distribution. However, Atomic cell estimates of the number of households differ considerably to those obtained under Curtis's model and IlocA imputation cells. This suggest employing imputation based on Atomic cells is likely to give biased estimates. It is therefore less robust in the upper tail of the energy value distribution. When these results are considered in the round, it appears IlocA reproduces much of the two-way main effects model estimates found by Curtis et. al. (2015) in marginal terms, with some key differences. These differences being sufficient to justify caution on the part of the policymaker or homeowner, as the potential cost of an energy retrofit in Ireland can be up to €75,000 per dwelling in 2022.

## 8. Closing Remarks

In this study, we set out a straightforward and novel bottom-up procedure called IlocA. It creates imputation cells based on frequency counts in a two-way contingency table. This table is generated from a pair of fully observed survey dataset classification variables. The cross-classification of these two variables partition the dataset into mutually exclusive and exhaustive cells, called atomic cells. IlocA aggregates these cells into larger clusters/aggregation groups that we call imputation cells. Imputation cells created by IloA are designed to have a minimum cell frequency count, to ensure imputed mean estimates will be statistically efficient and robust. IlocA also only groups together cells associated with near-zero log-odds ratios. These ratios define dependence or association between the classification variables. Thus, IlocA also ensures dependency between the classification variables is not attenuated. This approach to creating imputation cells is novel. Moreover, IlocA has been purposely conceived with a view to creating imputation cells where one classification variable is geographic. We highlighted, these so called geographic contingency tables, tend to have a predominance of independent log-odds ratios and so tend to generate an under-dispersed log-odds ratio distribution. Importantly, in section 2 we justified this conjecture, thereby providing a theoretical basis for IlocA. Meanwhile, in section 3 we set out and discussed the salient features of the IlocA algorithm. When used for creating imputation cells, one nice additional feature of the procedure is it forms aggregation groups homogeneous with respect to atomic cell response means.

We conducted a set of simulation studies to to test the efficacy of IlocA. First, we examined IlocA in a pure contingency table cell clustering Our simulations showed IlocA *groups independent cells only* in a consistent and credible way, suggesting the procedure controls for type 1 error. In sections 5 and 6 we conducted a series of simulations to verify the consistency and credibility of the procedure under various missingness scenarios. Our focus was primarily on validating the quality of imputed overall dataset mean estimates generated when IlocA's imputation cells were deployed. We used cell mean imputation throughout to avoid conflating imputation method error with imputation cell generation error, as out interest was on the latter. Under eight different imputation settings, based on different data generating and response models, we showed IlocA's generated between twenty and twenty-five imputation cells. This figure being close to the optimal number identified based on exact response variable quantiles. For ignorable non-response the relative bias and relative root mean square error values compared favourably with ideal values obtained from deterministic regression of response variable quantiles. With non-ignorable missingness results were less favourable. Nonetheless, results were shown be consistent with those obtained elsewhere – this is comforting.

Finally, in section7, we deployed IlocA to impute missing building energy values. We adopted a geographic classification variable and a classification variable based on year of construction and dwelling heating type. We created a cross-classification contingency table based on these two variables and fed it to IlocA to generate imputation cells. Our results showed the imputation cells generated included several well-defined large groups of aggregation cells based in specific row or column of the contingency table. This indicated the two-way main effect model that had been adopted elsewhere to model this data had merit. Nonetheless, it was also evident the main effects approach was inadequate where two-way interactions were more appropriate. In this situation imputed estimates of households obtained by from IlocA differed significantly from previous estimates. This finding is important in a policymaking context, as retrofitting cost in the subgroup of dwellings considered will be considerably underestimated if based on previous estimates.

**Appendix**

In the following assertion, we characterise the density of the log-odds ratio of a blocked contingency table in terms of three component densities. Accordingly, this is not an assertion about the density of the sum of three random variables.

<u>Assertion</u>: For a large contingency table $A$ comprising two blocks $A_L$ and $A_R$, with column dimensions $n_L$ and $n_R$ respectively, and with $n_L \gg n_R$ and $a_{ij,R} \gg a_{ij,L}$, the asymptotic density of the log-odds ratios $\phi_{ijkl}$ of $A$, has dominant central peak originating from sub-table $A_L$.

<u>Verification</u>: For each sufficiently large sub-table $A_L$ and $A_R$, asymptotic normality of the log table values implies

$$\begin{aligned}\log a_{ij,L} &\sim N(0,\sigma_L^2)\\ \log a_{ij,R} &\sim N(0,\sigma_R^2)\end{aligned} \Rightarrow \begin{aligned}\phi_{ijkl,L} &\sim N(0,4\sigma_L^2)\\ \phi_{ijkl,R} &\sim N(0,4\sigma_R^2)\end{aligned}$$

Further, given $a_{ij,R} \gg a_{ij,L}$ and asymptotic normality, then $\sigma_R^2 \gg \sigma_L^2$.

Manifestly, each sub-table has $m(m-1) \times n_L(n_L-1)/4$ and $m(m-1) \times n_R(n_R-1)/4$ log-odds ratios $\phi_{ijkl,L}$ and $\phi_{ijkl,R}$ respectively. However, there is an intersection to consider and here there are $m(m-1)\, n_L n_R/2$ log-odds ratios $\phi_{ijkl,L\cap R}$. The distribution of these is asymptoticly normal, with variance in proportion to relative dimensions of sub-table blocks

$$\phi_{ijkl,L\cap R} \sim N\left(0, \frac{4}{n}\{n_L\sigma_L^2 + n_R\sigma_R^2\}\right)$$

Now, since $n_L \approx n$, the distribution of the log-odds ratios $\phi_{ijkl}$ associated with the full table $A$, is dominated by the $m(m-1) \times n_L(n_L-1)/4$ log-odds ratios $\phi_{ijkl,L}$ coming from $A_L$. Furthermore, since their variance $\sigma_L^2$ is relatively small, they are relatively more concentrated near zero. In relative terms, the size of the sets $\phi_{ijkl,L}$, $\phi_{ijkl,L\cap R}$ and $\phi_{ijkl,R}$ are $O(n^2)$, $O(n)$ and $O(1)$ respectively. Thus, there are orders of magnitude fewer log-odds ratios $\phi_{ijkl,R}$ and $\phi_{ijkl,L\cap R}$ and their respective variances are much larger than $\sigma_L^2$. By simply combining these three sets $\phi_{ijkl,L}$, $\phi_{ijkl,L\cap R}$ and $\phi_{ijkl,R}$, we get a mixture density of their respective log-odds ratios, which is also the density of the log-odds for the full table $A$. Consequently, the asymptotic density of the log-odds ratios $\phi_{ijkl}$ for the full table $A$, has a dominant central peak associated with sub-table $A_L$.

## SAS code to implement the procedure

```sas
* Test program to get log cross product in KASS 1980 Simulation
dataset - correct version working on frequency table ;

options orientation=landscape nocenter papersize=A4 linesize=132
pageno=1;
%inc "\\SASFILE\Environment\autoexec.sas";

%macro color ;

%do ii = 1 %to 1 ;

proc iml ;
  start OddsRatio (B0, M1, S1, cp, ijkl, bijkl, mijkl, sijkl,
altman, altmanV, OK) ;
  B = B0 ;
  m=nrow(B) ; n = ncol(B) ;   s=0 ;
  do i = 1 to m-1 ;
  do k = i+1 to m ;
  do j = 1 to n-1 ;
     do l = j+1 to n ;
       if (B[i,j] > 0) & (B[k,l] > 0 ) & ( B[k,j] > 0 ) & (B[i,l] >
0 ) then
       do ;
             odds = (B[i,j]*B[k,l]) / ( B[k,j]*B[i,l] ) ;
                lods = log(odds) ;
                alods = abs(lods) ;
                cp = cp // ( odds || lods || alods ) ;
                ijkl = ijkl // (i || j || k || l) ;
                bijkl = bijkl // ( B[i,j] || B[k,j] || B[i,l] ||
B[k,l] ) ;
                mijkl = mijkl // ( M1[i,j] || M1[k,j] || M1[i,l]
|| M1[k,l] ) ;
                sijkl = sijkl // ( S1[i,j] || S1[k,j] || S1[i,l]
|| S1[k,l] ) ;
       end ;
       end;
  end ;
  end;
  end;

  *print cp;
  altman=0; altmanV=0; OK=0;
  if nrow(cp) > 0 then
  do;
    altman = ssq(cp[,2]) ; *print altman ;
    nn = (1:nrow(cp) )`;
    altmanV = altman/nrow(cp) ; *print altmanV ;
    OK=1;
  end ;
  finish ;

  start ichisq ( A, chisq, chival, dof ) ;
```

```
    IA = A;
     tA=0;
    do i = 1 to nrow(A) ;
    do j = 1 to ncol(A) ;
       if ( A[i,j] < 1.0E200 ) then tA = tA + A[i,j] ;
    end ;
    end ;
    do i = 1 to nrow(A) ;
        t0=0 ;
      do j = 1 to ncol(A) ;  if ( A[i,j] < 1.0E200 ) then t0 = t0 + A[i,j];end ;
        rA = rA // t0 ;
    end ;
    do j = 1 to ncol(A) ;
       t0=0 ;
       do i = 1 to nrow(A) ; if ( A[i,j] < 1.0E200 ) then t0 = t0 + A[i,j];end ;
        cA = cA // t0 ;
    end ;

    do i = 1 to nrow(A) ;
    do j = 1 to ncol(A) ;
      if ( A[i,j] < 1.0E200 ) then IA[i,j] = rA[i]*cA[j]/tA;
    end ;
    end ;
 * Print IA [format=4.2];
    chisq = 0 ;
    do i = 1 to nrow(A) ;
    do j = 1 to ncol(A) ;
      if ( A[i,j] < 1.0E200 ) then
      chisq = chisq + ( (A[i,j] - IA[i,j])**2/IA[i,j] ) ;
    end ;
    end ;
    dof = (nrow(A)-1)*(ncol(A)-1);
    chival=cdf('CHISQ',chisq,dof) ;
   * print chisq chival dof;
   finish ;

 ************************************************************ ;
   start OddsRatio (B0, cp, ijkl, bijkl, altman, altmanV, OK) ;

   start OddsRatio (B0, M1, S1, cp, ijkl, bijkl, mijkl, sijkl, altman, altmanV, OK) ;

   call streaminit(&ii);
   maxsteps=20 ;
   rdim=8;
   cdim=5;
   A = repeat(0,rdim,cdim);
   vA = repeat(0, 480, 3);
   do t = 1 to 480 ;
      u = rand("uniform");
      i = mod(t,rdim) ; if i = 0 then i = rdim ;
       j = int(1+cdim*u) ;
       A[i,j]=A[i,j]+u;
```

```
      vA[t,1]=i;
      vA[t,2]=j;
      vA[t,3]=u;
   end ;

   print A [format=4.3];
   tA = sum(A) ;
   print " ************************************ &ii **************
" tA ;
   *print vA ;
   mdesign = design(vA[,1]);
   ndesign = design(vA[,2]);
   mcols = unique(vA[,1]);
   ncols = unique(vA[,2]);
   *print mdesign[colname=mcols] ndesign[colname=ncols];
   *print mcols ncols;
   A1 = repeat(0,ncol(mcols), ncol(ncols) );
   do i = 1 to ncol(mdesign) ;
     do j = 1 to ncol(ndesign) ;
        A1[i,j] = mdesign[,i]`* ndesign[,j] ;
     end ;
   end ;
   print A1[rowname=mcols colname=ncols];
   print " ***************************************************** " ;

   A = A1;

   B = A ;

   call ichisq ( A, chisq, chival, dof) ;
   call symputx('CHISQ', chisq) ;
   call symputx('CHIVAL', chival) ;
   call symputx('DOF', dof) ;
   call OddsRatio (B, cp, ijkl, bijkl, altman, altmanV, OK) ;
   print altman altmanV OK ;
   AI = &ii || altman ;

   Bmin=min(B) ;

   Bmin=max(B) ;

   isteps_stop=0;Tcells=rdim*cdim;
   do isteps = 1 to maxsteps ;
*print "******* " isteps;
   call OddsRatio (B, cp, ijkl, bijkl, altman, altmanV, OK) ;
   print altman altmanV OK ;
 * print cp ;
   if ( OK = 1 ) then
   do;
   cp = cp || ijkl || bijkl ;
   call sort(cp,{3}) ;
   if ( isteps=1) then cp0=cp;
   penalty = Tcells/ncol(loc(B>0));*/Tcells ;
   print isteps penalty;

   MinCellVal = min(cp[1,8:11]) ;
```

```
   print MinCellVal ;

   OKc=0;
      if ( OKc = 0 ) then    do ; cellrow = cp[1,6] ; cellcol = cp[1,4] ; if ( B[cellrow,cellcol] = MinCellVal ) then OKc=1 ; end ;
      if ( OKc = 0 ) then    do ; cellrow = cp[1,6] ; cellcol = cp[1,5] ; if ( B[cellrow,cellcol] = MinCellVal ) then OKc=1 ; end ;
      if ( OKc = 0 ) then    do ; cellrow = cp[1,7] ; cellcol = cp[1,4] ; if ( B[cellrow,cellcol] = MinCellVal ) then OKc=1 ; end ;
      if ( OKc = 0 ) then    do ; cellrow = cp[1,7] ; cellcol = cp[1,5] ; if ( B[cellrow,cellcol] = MinCellVal ) then OKc=1 ; end ;

 *  print  cellrow cellcol ;
      OK=0; altman1 = 0; altmanV1=0;
      call XOddsRatio (B, cellrow, cellcol, tcp, altman1, altmanV1, OK) ;

 *    print " ************* OK = "  OK  altman1 altmanV1 ;
   do i = 1 to nrow(tcp) ;
   mdif = 1.0e200 ;
      do j = 9 to 11 ;
         dif = abs(tcp[i,j]-tcp[i,8]);
         if dif < mdif then do ; mdif = dif ; val = tcp[i,j] ; end ;
       end ;
      MCV = MCV // (val || mdif ) ;
   end ;
   tcp = tcp || MCV ;
   call sort(tcp,{13,3}) ;
*     print tcp ;
   cummin=tcp[1,12];
   do i = 2 to nrow(tcp) ;
     cummin = cummin + tcp[i,12] ;
       if cummin > 20*penalty then do ; istop=i; i=nrow(tcp)+1; end;
   end ;
   if cummin <= 20*penalty then istop = nrow(tcp) ;
*  print istop cummin;
   tcp = tcp[1:istop,];
*     print tcp ;

   rc = -1*isteps || cellrow || cellcol ;
   do i = 1 to nrow(tcp) ;
      if ( tcp[i,9]  = tcp[i,12] ) then do ; r = cellrow  ; c = tcp[i,7] ; end ;
      if ( tcp[i,10] = tcp[i,12] ) then do ; r = tcp[i,6] ; c = cellcol    ; end ;
      if ( tcp[i,11] = tcp[i,12] ) then do ; r = tcp[i,6] ; c = tcp[i,7] ; end ;
      rc = rc // ( -1*isteps || r || c ) ;
   end ;
*     print rc ;

   do i = 1 to nrow(rc) ;
      B[rc[i,2], rc[i,3]]=-1*isteps;
   end ;
   free cp ijkl bijkl tcp MCV;
*  print B;
```

```
    call OddsRatio (B, cp, ijkl, bijkl, altman, altmanV, OK) ;
    print isteps altman altmanV ;

    rcmap = rcmap // rc ;
    free rc ;
    do i = 1 to nrow(B) ; do j = 1 to ncol(B) ; if ( B[i,j] > 0 ) &
(B[i,j] < Bmin ) then Bmin = B[i,j] ; end ; end;
    if (Bmin > 20*penalty ) then  do ; isteps_stop=isteps;
isteps=1.0e300; end;
    end ;
    else do ; isteps_stop=isteps; isteps=1.0e300; end;
    end ;

    print isteps isteps_stop Bmin ;
 * print rcmap ;

    do j = 1 to ncol(B) ;
      ux = t(unique(B[,j])) ;
        do k = 1 to nrow(ux) ;
          ss=0;
          do l = 1 to nrow(B) ;
            if (ux[k]=B[l,j]) then ss = ss+1;
          end;
          vx = vx // ss ;
        end ;
        vx = ux || vx ;
      uB = uB // vx ;
        free vx ux;
    end ;
/*
  * assign few stragglers to lowest assignment group in that
row/column;
    do r = 1 to nrow(B) ; do c = 1 to ncol(B) ;
      if ( B[r,c] > 0 & B[r,c] < 20 ) then
      do;
        maxneg = -10 * MAXSTEPS ;
        print r c maxneg ;
        do i = 1 to nrow(B) ;
          if ( B[i,c] > maxneg) & ( B[i,c] < 0 )  then  maxneg = B[i,c]
;
      end ;
        do j = 1 to col(B) ;
          if ( B[r,j] > maxneg) & ( B[r,j] < 0 )  then maxneg = B[r,j]
;
      end ;
      B[r,c] = maxneg ;
        end ;
    end ; end ;
*/

    isteps = isteps_stop ;
    do r = 1 to nrow(B) ; do c = 1 to ncol(B) ;
      if ( B[r,c] > 0 & B[r,c] < 20 ) then
      do;
        isteps=isteps+1 ;
```

```
        cellrow=r ; cellcol=c ;
        OK=0; altman1 = 0; altmanV1=0;
        call XOddsRatio (B, cellrow, cellcol, tcp, altman1, altmanV1,
OK) ;
         if ( OK = 1 ) then
         do ;
         do i = 1 to nrow(tcp) ;
           mdif = 1.0e200 ;
           do j = 9 to 11 ;
             dif = abs(tcp[i,j]-tcp[i,8]);
             if dif < mdif then do ; mdif = dif ; val = tcp[i,j] ;
end ;
            end ;
            MCV = MCV // (val || mdif ) ;
         end ;
         tcp = tcp || MCV ;
         call sort(tcp,{13,3}) ;
         cummin=B[cellrow,cellcol];
         do i = 1 to nrow(tcp) ;
             cummin = cummin + tcp[i,12] ;
          if cummin > 20 then do ; istop=i; i=nrow(tcp)+1; end;
         end ;
         if cummin <= 20 then istop = nrow(tcp) ;
*  print istop cummin;
         tcp = tcp[1:istop,];

         rc = -1*isteps || cellrow || cellcol ;
         do i = 1 to nrow(tcp) ;
             if ( tcp[i,9]  = tcp[i,12] ) then do ; r = cellrow  ; c =
tcp[i,7] ; end ;
             if ( tcp[i,10] = tcp[i,12] ) then do ; r = tcp[i,6] ; c =
cellcol    ; end ;
             if ( tcp[i,11] = tcp[i,12] ) then do ; r = tcp[i,6] ; c =
tcp[i,7] ; end ;
             rc = rc // ( -1*isteps || r || c ) ;
          end ;

        do i = 1 to nrow(rc) ;
           B[rc[i,2], rc[i,3]]=-1*isteps;
        end ;
        free cp ijkl bijkl tcp MCV;
        call OddsRatio (B, cp, ijkl, bijkl, altman, altmanV, OK) ;
        rcmap = rcmap // rc ;
         free rc ;
       end ;
      end ;
  end ; end ;
/*
  * assign few stragglers to lowest assignment group in that
row/column;
  nvals = 0 ;
  do r = 1 to nrow(B) ; do c = 1 to ncol(B) ;
    if ( B[r,c] > 0 & B[r,c] < 20 ) then
    do;
      maxneg = -10 * MAXSTEPS ;
      if ( rand("uniform") < 0.5 ) then
```

```
      do ;
      do i = 1 to nrow(B) ;
        if ( B[i,c] > maxneg) & ( B[i,c] < 0 )  then  maxneg = B[i,c]
;
   end ;
     end ;
      else
      do ;
      do j = 1 to ncol(B) ;
        if ( B[r,j] > maxneg) & ( B[r,j] < 0 )  then maxneg = B[r,j]
;
   end ;
     end ;

   B[r,c] = maxneg ;
   rcmap = rcmap // ( maxneg || r || c) ;
     end ;
  end ; end ;
*/
  C = A ; *repeat(0,rdim,cdim);
  minB = min(B) ;
  do k = -1 to minB by -1 ;
     tt=loc(B=k) ;
       if ( ncol(tt) > 0 ) then do ;
     bb = mean(A[tt]) ;
     C[tt] = bb;
       end ;
       free tt bb ;
  end ;
  call OddsRatio (C, cp, ijkl, bijkl, altman, altmanV, OK) ;
  print isteps altman altmanV ;
  AI = AI || altman ;

  print  B ;
 * print isteps isteps_stop Bmin ;
 * print rcmap ;

  if nrow(cp0) > 0 then do ;
    create dcp0 from cp0 ;
    append from cp0 ;
  end ;

  create uB from uB ; append from uB;
  create tAI from AI ; append from AI;

quit ;
run ;
proc summary data=uB nway;
class col1 ;
var col2 ;
output out=suB sum= ;
run ;
data suB (keep=color col2) ;
  set sUB ;
  if col1 < 0 ;
```

```sas
   color=-1*col1 ;
run ;
proc sort data=suB ; by color ; run;
proc transpose data=sub out=tub prefix=color ;
id color;
var col2 ;
run ;

data tuB ; set tuB ; sim = &ii ; chisq=round(&CHISQ,0.001) ;
chiProb=round(&CHIVAL,0.001); dof=&DOF; run;
proc sort data=tuB ; by sim ; run ;
proc sort data=tAI ; by col1 ; run ;

%if ( &ii = 1 ) %then %str( data color ; set tuB ; run ; ) ; %else
%str( data color ; merge color tuB ; by sim ; run ;) ;
%if ( &ii = 1 ) %then %str( data AI ; set tAI ; run ; ) ; %else
%str( data AI ; merge AI tAI ; by col1 ; run ;) ;

%end ;

%mend ;

%color ;
*****************************************************************
*********** ;
*****************************************************************
**************** ;
quit ;
```